\documentclass{emulateapj}
\usepackage{graphicx}
\usepackage{amsmath}
\usepackage{afterpage}
\usepackage{color}
\usepackage{amsmath}
\def\resp{respectively}

\def\Msun{{\rm M_\odot}}

\pdfoutput=1
\slugcomment{(Draft \today)}

\gdef\ltsima{$\scriptscriptstyle \; \buildrel < \over \sim \;$}
\gdef\simlt{\lower.3ex\hbox{\ltsima}}

\gdef\gtsima{$\scriptscriptstyle \; \buildrel > \over \sim \;$}
\gdef\simgt{\lower.3ex\hbox{\gtsima}}

\gdef\about{\raise.3ex\hbox{$\scriptscriptstyle \sim $}}

\def\sigbar{\overline\sigma}

\shortauthors{Kelson, Benson, \& Abramson}

\shorttitle{On the Origin of the Galaxy Stellar Mass Function}

\begin{document}

\title{On the Origin and Evolution of the Galaxy Stellar Mass Function}

\author{Daniel D.~Kelson, Andrew J.~Benson}
\affil{The Observatories, The Carnegie Institution for Science,\\
813 Santa Barbara St., Pasadena, CA 91101,\\
@grumpykelson, @abensonca}

\and

\author{Louis E.~Abramson}
\affil{Department of Physics and Astronomy, University of California,\\
430 Portola Plaza, Los Angeles, CA 90095,\\
@lab\_rams}

\begin{abstract}
Here we explore the evolution of galaxy ensembles at early times by writing the {\it in situ\/} stellar mass growth of galaxies purely as a stationary stochastic (e.g.,
quasi-steady state) process. By combining the mathematics of such processes with Newtonian gravity and a mean local star formation efficiency, we show that the stellar mass
evolution of galaxy ensembles is directly related to the average acceleration of baryons onto dark matter halos at the onset of star formation, with explicit dependencies on
initial local matter densities and halo mass. The density term specifically implies more rapid average rates of growth in higher density regions of the universe compared to low
density regions, i.e., assembly bias. With this framework, using standard cosmological parameters, a mean star formation efficiency derived by other authors, and knowledge of the
shape of the cosmological matter power spectrum at small scales, we analytically derive (1) the characteristic stellar masses of galaxies ($M^{*}$), (2) the power-law low-mass
slope ($\alpha$) and normalization ($\phi^{*}$) of the stellar mass function, and (3) the evolution of the stellar mass function in time over $12.5\geq z\gtrsim 2$.
Correspondingly, the rise in the cosmic star formation rate density over these epochs, while the universe can sustain unabated fueling of star formation, also emerges naturally.
All of our findings are consistent with the deepest available data, including the expectation of $\alpha\approx -7/5$; i.e., a stellar mass function low-mass slope that is notably
shallower than that of the halo mass function, and with no systematic deviations from a mean star formation efficiency with density or mass, nor any explicit, additional feedback
mechanisms. These derivations yield a compelling richness and complexity but also show that very few astrophysical details are required to understand the evolution of cosmic
ensemble of galaxies at early times.
\end{abstract}




\section{Introduction}

Stories about the formation of the Milky Way have been told for thousands of years, yet our ability to comprehend the
origins of our Galaxy, as well as all the other ones, remains limited. Today tales of galaxy formation are spun
using ingredients from a modern cosmographical context, based on
gravitational collapse \citep{peebles1967}, on the deposition of baryons onto dark matter halos \citep{white1978},
on the hierarchical growth of structure through cosmic time \citep{blumenthal1984},
and informed by the long-term effects of early density fluctuations on the physical distributions of galaxy types \citep{dressler1980}.
Numerical experiments that combine these basic tenets, {\it N}-body and hydrodynamical simulations, rapidly became the primary mechanisms with
which to theorize all we would strive to know about the Milky Way's formation and the growth of all galaxies
\citep[e.g.][]{vogelsberger2014,hopkins2014,mcalpine2016}.

And while these computational efforts have helped elucidate what must inherently be a
complicated picture behind galaxy formation, there has been the unnatural side effect of
rendering opaque to the general community the physical processes (and their parameters) that dominate how and why galaxies became what they are.
In other words, the progression of technologies and capabilities of simulation science has separated many astronomers from the core
astrophysical principles that underpin modern visions of galaxy formation. So while
simulations now provide detailed pictures, many confrontations of data with theory are executed with
relatively simple pictures and models \citep{dekel2006,dutton2010,dave2012},
with each turning on dramatic, life-changing events such as ``quenching'' \citep[meaning different things
to different people;][]{lin1992,lilly2013,pacifici2013,abramson2016,pacifici2016}
in order to provide explanations for why galaxy colors come in red and blue, why some galaxies appear simply to have died
\citep{whipple1935,humason1936,tinsley1968}. 

Certainly, existing data may not support {\it any\/} toy models uniquely \citep{abramson2016}, even though
an empirical picture that describes the cosmic evolution of galaxy ensembles is particularly desirable. Having an empirical
picture, or at least one grounded solely in the data, would free the community from overly
rigid assumptions, including those that underpin halo occupation distribution models or abundance matching
\citep[][and many others]{berlind2002,kravtsov2004,vale2004,guo2010}.
However, deriving an empirical picture within the framework of ($\Lambda$)CDM
has so far not easily allowed for significant decoupling between stellar mass growth and halo growth, as implicit
equivalence between the two is a major, unproven assumption in standard treatments of galaxy evolution \cite[e.g.][]{behroozi2013,gu2016}.
At best it is assumed that the diversity of DM assembly fully encapsulates the diversity in galaxy growth, but, again, the assumption comes
with many caveats, remaining undemonstrated in its validity, and demonstrably invalid at late times,
such as when galaxy groups grow through
continued hierarchical assembly \citep{williams2012} while their constituent galaxies grow more slowly \citep{patel2011}, and while
baryons retain the ability to radiatively cool with respect to the dark matter \citep[e.g.][]{wetzel2015}.

Furthermore, empirical derivations of galaxy formation and assembly have also treated the mean
evolution of ensembles of galaxies as representative of what galaxies actually do \citep{peng2010,leitner2012,pvd2013,papovich2015},
an assumption that does little to help illuminate the true underlying distribution of growth histories, to say nothing of those of
individual galaxies \citep[see, e.g.,][]{pacifici2013,gladders2013,kelson2014,abramson2015,pacifici2016}.

Part of the problem is that galaxy ensembles show so little evolution in the mean over the past several Gyr, at fixed
mass or number density, that it seems natural to translate this slow evolution in bulk properties \citep[e.g. in color, in $M/L$ ratio, in morphology][]{koo1992,kelson1997,bell2004,vdw2007}
into pictures in which the evolution of individual galaxies is equally slow or static
\citep{leitner2012}, at least until it suddenly isn't \cite[e.g.][]{peng2010,peng2012}.
In such pictures, specific intrinsic and extrinsic properties determine the fates of systems today through dramatic
and permanent {\it transformations\/}. In other words, galaxy evolution and quenching are often discussed
in terms analogous to how businesses go bankrupt:  gradually, and then suddenly.

Even discussions of ``fast-track'' and ``slow-track''
quenching may be distinctions without a difference, as the time for galaxies cross from blue to red is always a fixed fraction of a Hubble
time \citep{abramson2016,zolotov2015}. And if that means stellar population indicators provide very little information about galaxy formation
timescales, the nearly 1:1 correlation between galaxy star formation rates and stellar masses
\citep{brinchmann2004,noeske2007,whitaker2014} also suggests that the processes of galaxy formation and assembly may also be largely scale-free \citep{kelson2014}.

And while simple models help to give insight into how and why galaxies look they way they do, \citet{tinsley1968} remains a stern
warning to this day against overly deterministic formation scenarios. Such work demonstrated the potential, and potentially hidden, ubiquity
of diversity in star formation histories. Together with \citet{tinsley1978}'s positing of (relative) balance between the forces that make
stars (gravity), the forces that bring fuel to make those stars (gravity), and the forces that stars re-impose on their environments
(feedback), this might already have led us to a number of conclusions about the evolution and behavior of galaxy ensembles without adding many
additional mechanisms. By now this notion that stellar mass growth occurs in states of quasi-static equilibrium,
between the inflows that fuel star formation and the resulting outflows has been explored in great detail
\citep[e.g.][and others]{efstathiou2000,dave2012,lilly2013,dekel2014}. However, the extent to which
diversity of growth histories and (quasi-) steady state equilibrium go hand-in-hand \citep{efstathiou2000} remains under-appreciated.

The work here is an attempt to re-frame what average growth histories mean, and to change the way we interpret core statistical metrics of
galaxy evolution. We begin construction of a new theoretical framework for thinking about the evolution of cosmological galaxy ensembles.
Until now, there have principally been two of these: purely numerical simulations \citep[e.g.][and many since]{bertschinger1998} and semi-analytical schemes
built on merger trees \citep[e.g.][and many since]{kauffmann1993,cole1994}.

Our derivations lie in a third vein, more aligned with a statistical mechanical approach. By positing that stochasticity and stationarity are basic principles
underlying halo and galaxy growth (Section \ref{sec:terms}), we are able to retain an analytic framework for describing the long-term behavior of
ensembles, while avoiding a reliance on mean-based parametric prescriptions requiring fine-tuning. Such an approach incorporates the dynamism of the
(our) cosmological system, not only jettisoning deterministic histories for individual members of the ensemble, but explicitly refutes the notion that
individual, deterministic histories may be written. By doing so, we are able to elucidate the full extent to which very basic phenomena (e.g., the
matter power spectrum) {\it dictate\/} the evolution of the full suite of early galaxy populations so as to define the emergence and behavior of the stellar
mass function {\it absent, effectively, any additional astrophysics\/}.

The goal of this work, then, is to show that galaxy evolution can be thought of as the stochastic, cosmological, jostling of ensembles of quasi-steady
state systems\footnote{Or at least ensembles of quasi-queuing systems. While stationarity naturally arises out of discussions of steady state equilibrium (see
references given earlier), the condition of steady-state is not technically required for us to represent stellar mass growth as a
stationary stochastic process. The transition of material out of the cosmological context and into queues for conversion to stars, galaxies, on cosmologically
short timescales implies stationarity through Little's Law \cite[see][]{little2008,whitt1991}. In other words, even when the galaxies themselves are in unsteady states
the baryons spend time in the queue awaiting their big moments to be turned into stellar mass.}.
If so, we can bring to bear a wealth of mathematical tools developed to describe the evolving ensembles \citep{kolmogorov1940,hurst1951,mandelbrot1968,koutsoyiannis2002}.
We demonstrate the power of this approach here by inferring the form and evolution of the
galaxy stellar mass function over the first few billion years of cosmic time (when fuel supplies are ample), but presumably far more can be examined in this formalism.
As a side effect, this derivation---which seeks the ``equations of motion'' for galaxy ensembles --- will illuminate the critical role played by initial conditions, revealing how
the matter power spectrum itself naturally imposes correlations between stochastic changes that ultimately dictates the shape and normalization of the
stellar mass function without assumptions about sources of feedback (e.g., supernovae or active galactic nuclei).

In what follows, we adopt a Hubble constant, $H_0=68\, {\rm km\, s^{-1}\, Mpc^{-1}}$; a mean matter density of the universe,
$\Omega_M=0.31$; a cosmological constant, $\Omega_\Lambda=0.69$; a mean baryon fraction,
$f_b=0.16$ \citep{planck2015}; and a mean efficiency for the conversion of baryons to stars, $\epsilon=0.015$ \citep{krumholz2005}.

In Section \ref{sec:terms} we define a few important concepts, followed by
Section \ref{sec:stoch} with a brief review of ({\it in situ\/}) mass growth as a stationary stochastic process. In Section \ref{sec:sigbar} we derive
the normalizations of the stochastic growth histories from the characteristic acceleration of baryons onto the seeds of galaxies at $z\sim
10-15$. These ingredients are brought together in Section \ref{sec:mfs} to derive purely theoretical stellar mass functions that are then compared
to data from the literature. Finally, our conclusions are summarized in Section \ref{sec:disc}.

\section{Terminology}
\label{sec:terms}

In this work we employ a number of terms, some familiar to the reader, but that have specific technical definitions that
may run counter to one's intuition.

\subsection{``Stochastic Process''}
\label{sec:stochDef}

The word ``stochastic'' has often been used in the astronomical literature as synonymous with ``bursty'' or ``random.''
Mathematically, however, one should think of a stochastic process as very nearly the opposite. Some intuition into
stochastic processes can be developed by recalling the fundamental theorem of calculus.
Starting with an arbitrary continuous function $f(x)$, one approximates its derivative by calculating
$\left[f(x+h)-f(x)\right]/h$ for successively smaller values of $h$, generating the derivative in the limit $h\rightarrow 0$.
In our case, however, one can think of a galaxy's star formation history as $f$, writing down the star formation rates at times $t$ and $t+\Delta t$ as $f(t)$
and $f(t+\Delta t)$.

While your calculus instructor would have taken the limit as $\Delta t\rightarrow 0$, we make no such
requirement, and instead allow $\Delta t$ to remain arbitrarily large. As such, we allow $f(t)$, or ${\dot M}(t)$
in our case, to be non-differentiable (yet integrable). The stochastic process is thus a discontinuous function, having no definable derivative.

In order to make progress on this general class or set of functions, one
{\it replaces the property of continuity with a general description of the discontinuities\/}, e.g. by parameterizing the distribution of
their amplitudes.
(Instead of requiring the limit of $h\rightarrow0$, we require knowledge of how big the jumps we encounter by {\it not} enforcing continuity are likely to be.)
When they have a normal distribution, for example, these discontinuities are referred to as Gaussian noise; {\it the stochastic process itself is the
integral of that Gaussian noise, and is then otherwise known as Brownian motion\/}.
More mathematical descriptions and detail will be in the next Sections, but so long as
one can apply central limit theorems to the distributions of these discontinuities, one has access to the set of rules governing stochastic calculus:
with limit theorems, one can derive probabilities for $f$, and because these probability distributions are continuous in time, they are easy to work with.
For most purposes (including ours), one simply derives the first and second moments of those probability distributions
--- referred to as ``expectation values.''  These often end up being relatively simple functions of time, and are typically what can be compared to observations.
Fundamentally, in a stochastic system, the {\it ensembles behave deterministically}, though their constituents {\it do not}, a fact that
is critical to our derivations (Section \ref{sec:sigbar}).

This (standard) definition of a stochastic process stands in contrast to a common notion in astronomical literature that ``stochastic''
implies arbitrary levels of randomness, complete anarchy, or necessarily any level of general burstiness with periods of inactivity
punctuated by periods of intense activity.

\subsection{``Stationary''}

A ``stationary'' stochastic process is one in which, given a complete,
unbiased ensemble of histories, {\it the arithmetic mean\/} of those discontinuities, or stochastic changes, is identically zero at each instant of time.
In other words, for a given object experiencing a stochastic trajectory, the value of the process is expected not to change from one
timestep to the next. The {\it average\/} change, taking all objects in a fair ensemble, in fact remains zero, even though individual growth
histories will wander (a bit) from the values at the previous timestep.

One physical interpretation of this definition is that it is equivalent to quasi-steady state growth.
When fuel for the stochastic process of stellar mass growth remains abundant, galaxies can continue growing with inflows and outflows in rough
balance, in quasi-steady states, with no expectation for star formation rates to either go up or down, though they individually may.
Such an interpretation of stationarity is a natural way to connect earlier work
\citep{tinsley1978,efstathiou2000,dave2012,lilly2013,dekel2014}, but not uniquely so.

When the timescales for baryons to be converted to stars in galactic systems, or ensembles of galactic systems,
is mismatched with the timescales driving stochastic changes to galaxy growth one finds that
Little's Law holds \citep{little2008} strictly due to conservation of mass entering the ``queuing system'' that converts baryons to stars.
As a result, the stochastic process of stellar mass growth remains a stationary one even when not in steady state equilibrium (see Appendix
\ref{sec:app4}).

\subsection{``Markovian'' vs. ``Non-Markovian''}
\label{sec:markovian}

Additionally we can also write that the stochasticity is Markovian when the discontinuous jumps in the star formation history
for a given galaxy are uncorrelated over the lifetime of that galaxy. When this assumption is not valid, the stochastic
process is non-Markovian. 

To be clear, this simply means that the changes in an individual
galaxy's SFRs are correlated over its lifetime. The phenomenon of positive correlations
between stochastic changes (such that upwards ``bounces'' tend to breed further increases, not sharp declines) has long been
referred to as long-term memory \citep[e.g.][]{mandelbrot1968}, an attempt to describe how
systems can experience state changes that alter how later events may physically manifest (such as how
two different galaxies can experience the same late-time accretion events but because their earlier histories are different, the two systems
end up with physically different outcomes).

The spectrum of covariance between stochastic changes is often described by an additional parameter $H$---the ``Hurst parameter'' (see
Sections \ref{sec:meanevol} and \ref{sec:hurst}). Published data on the star-forming main sequence strongly imply a Hurst parameter of $H=1$
\citep{kelson2014}, implying maximal ``memory'' or longest-possible timescales of correlations between stochastic changes to
galaxy states over the lifetimes of galaxies.
Such a value for $H$ can arise naturally in stochastic flows \citep{kolmogorov1940,hurst1951,koutsoyiannis2002},
but it can also occur when there are simply large scale (or long timescale)
fluctuations in the fueling of the stochastic process \citep{samorodnitsky2006}, in this case galaxy growth.
These sorts of climatic changes to fueling of mass growth may be expected from a cosmological power spectrum with density fluctuations on all scales,
or at least scales that are relevant for galaxy evolution and structure formation.
This statement will become important when we construct and examine the emergence and evolution of the stellar mass function in Section \ref{sec:sigbar}.


\section{Early Mass Growth as a Stationary Stochastic Process}
\label{sec:stoch}


\subsection{Basic Mathematical Principles}

Here we review the mathematics that yields the long term behavior of ensembles of galaxies experiencing mass growth as a stationary
stochastic process \cite[see, e.g.][]{kelson2014}.
For this brief review of the math, we will assume a constant timestep $\Delta t$. We use uppercase $T$ to denote the integer index of a
time interval, and lowercase $t=T\Delta t$ as the elapsed time since the process began.

\subsubsection{Individual Galaxies and What We Need to Know About Them}

We will soon see that galaxies in our formalism are akin to gas particles in statistical mechanics: While, in principal, their individual
behaviors can be described (as, e.g., in SAMs), doing so is not necessary to forming an accurate theory of large numbers of them
that---critically---does not lose sight of their individual diversity. Here, to be able to move to such a powerful description, all we need is a
suitable {\it form\/} for the behavior of individual systems. It turns out that this form is quite general.

The rate of growth in stellar mass, $M$, for galaxy $i$ over the time interval from $t$ to $t+\Delta t$ is
\begin{equation}
{\dot M}_{i,t} \equiv \frac{M_{i,t+\Delta t}-M_{i,t}}{\Delta t}.
\label{eq:one}
\end{equation} 
Changes to these rates of growth can be expressed as $X$, such that:
\begin{equation}
	X_{i,T+1} = {\dot M}_{i,T+1}-{\dot M}_{i,T},
\label{eq:two}
\end{equation}
akin to a second derivative. Therefore, we have:
\begin{equation}
	{\dot M}_{i,T+1} = \sum_{T=0}^{T+1} X_{i,T}.
\label{eq:three}
\end{equation}
$X_{i,0}= 0$ can be assumed with no loss of generality.

So far, Equations \ref{eq:one}, \ref{eq:two}, and \ref{eq:three} are straightforwardly connected to a traditional view of SFHs:
they relate to the changes that actually occur from one time to another in an individual galaxy's SFR. Now, we are going to make a conceptual
leap. Rather than approximating the {\it individual} SFHs as deterministic, we allow them to conform to the general notion of stochastic as
``random'' processes (or, later, quasi-random ones, with embedded correlations).
As such, we will shortly lose track of the $i$-th galaxy---as one loses track of the $i$-th gas particle---but paying this price will grant us
access to an expansive description of the galaxy population.

\subsubsection{The Galaxy/Ensemble Jump and Crosstalk}

The basic formalism for individual galaxies above allows us now to model the distribution of accumulated changes across time for the entire galaxy population as per Section
\ref{sec:terms}.

There, we introduced the notion that we are interested in moments of probability distributions at various instances in time---the ``expectation values.''
Assuming there is a probability distribution for property $X$ for $i\in \{1\ldots N\}$ objects at time $T+1$, we would write ${\rm E\/}[X_{i,T+1}]$,
the first moment (mean) of the probability distribution of all $X_{i,t}$ (i.e., $\approx$ galaxy $\ddot{M}$) values at
time interval $T+1$, as
\begin{equation}
	{\rm E\/}[X_{i,T+1}] = \frac{1}{N}\sum_{i=1}^N X_{i,T+1}
\label{eq:stmean}
\end{equation}

The summation sign on the RHS of this equation signals that this is where the identities of individual galaxies drop out of the story.
Later, once one has derived how the ensembles evolve, representative SFHs for individual galaxies can be reconstructed, showing, e.g.,
how ``bumpy'' they are or are not \citep[see][]{kelson2014}.

Now we impose the requirement of stationarity. This demands that the mean (or sum) of the $X_{i}$ terms at fixed time in the RHS
in Equation \ref{eq:stmean} is {\it zero\/}:
\begin{equation}
	{\rm E\/}[X_{i,T+1}] = 0.
\label{eq:stationary}
\end{equation}
That is, the SFRs of individual galaxies are assumed {\it not to change} from one instant to the next (more below). If so, we now have the expectation
\begin{equation}
{\rm E\/}[{\dot M}_{i,T+1}] = {\dot M}_{i,T}
\label{eq:stationary2}
\end{equation}
That is, if ${\dot M}$ is a stationary process, the average value of ${\dot M}$ {\it for the ensemble} is also expected not to change
from one timestep to the next (Section \ref{sec:terms}). In other words, galaxy $i$, with its endless supply of fuel for star formation,
is {\it expected\/}, to have ${\dot M}_{i,T+1}={\dot M}_{i,T}$, but this is certainly not {\it required\/}.

This is because the distributions of $X_i$ have second moments. That is, galaxy $i$ can experience an event at time interval $T+1$ that
leads to a {\it nonzero} change in its star formation rate from ${\dot M}_{i,T}$ to ${\dot M}_{i,T+1}$.
This is stochasticity (again, see Section \ref{sec:terms}), and these changes occur with some variance
\begin{equation}
{\rm Var\/}[{\dot M}_{i,T+1}-{\dot M}_{i,T}] = {\rm Var\/}[X_{i,T}] \equiv \sigma^2_{i,T}
\label{eq:varX}
\end{equation}
As long as $X$'s domain is symmetric about zero, for a fair sample of systems, this will not violate Equation \ref{eq:stationary}.
The value $\sigma_{i,T}$ represents the width of the probability distribution of possible changes to the $i$-th galaxy's SFR at time interval $T$.
(Recall that a stochastic process is one characterized by the distribution of discontinuity amplitudes, as discussed in Section \ref{sec:terms}.)

Remarkably, the only requirement we place on these galaxy- and time-dependent variances, $\sigma^2_{i,T}$,
is that they be finite. {\it So far in this discussion we have no requirement that we have any physically meaningful estimates for these variances.\/}
But because these are bounded (i.e. never infinite) we can define a root-mean-square, or RMS, stochastic fluctuation over the
lifetime of changes to each galaxy's star formation rates up to time $t$:
\begin{equation}
{\overline\sigma}_{i,T} = \left({1\over T}\sum_{j=1}^T\sigma_{i,j}^2\right)^{1/2}
\label{eq:sigbar}
\end{equation}
This number, ${\overline\sigma}_{i,T}$, is thus directly related to the width of the probability distribution of star formation rates,
$P({\dot M}_{i,T})$, at time $t=T\Delta t$,
and, modulo a factor of $\sqrt{T}$, informs the average ``bumpiness'' of individual galaxy SFHs (prior to the epoch of observation).

Equation \ref{eq:sigbar} implies that there are numerous paths to achieving the same value of $\overline\sigma_{T}$, i.e. there
are numerous ways for different growth trajectories to be characterized by the same mean variance.
This is somewhat key: it both allows us to (1) define an ensemble of systems that achieve the same RMS stochastic fluctuation at time $t$ (or $T$)
and so can be treated probabilistically, and also (2) to generalize our findings to any other such ensemble, independent of the actual values of $\sigbar$.
Let us then refer to such an ensemble as $I$, whose defining common RMS stochastic fluctuation is ${\overline\sigma}_{I,T}$.


\subsection{How the Ensemble Expectations for Such ${\dot M}$ and $M$ Evolve with Time}
\label{sec:meanevol}

For the galaxies in ensemble $I$, we can derive probability distributions for star formation rates ($\dot{M}$) and stellar masses ($M$), normalized by
that RMS fluctuation, ${\overline\sigma}_{I,T}$.
If a given stochastic change to a galaxy's SFR is uncorrelated with every other change to its SFR, then,
because SFRs cannot be negative, the distribution of $\dot{M}_{I,T}$ is equal to the nonnegative side of a Gaussian.\footnote{This result follows from the
martingale central limit theorem and Doob's principle \citep{hall1980}.
Recall, it is the distribution of $X\approx\delta t\ddot{M}$ that must be symmetric about zero (Equation \ref{eq:varX}). But when
the boundary value of ${\dot M}\ge 0$ is enforced during the summation in Equation \ref{eq:three}, $X$ becomes a submartingale.}
But if the stochastic change to the SFR at time $t$ {\it is} correlated with those at
earlier times $t-m\Delta t$ ($0<m<\infty$), then that probability distribution will follow a different form.

Fortunately, one can still derive the first and second moments of $P({\dot M}_{I,T})$ analytically, with an extra $T^{H-1/2}$
going into Equation \ref{eq:sigbar} (see Equation \ref{eq:SFR}). Here $H$ is known as the Hurst parameter \citep[][see Section \ref{sec:hurst}, below]{mandelbrot1968},
and it encodes the spectrum of correlations between $X_{i,1}$, $X_{i,2}$, $X_{i,3}$, $\ldots$, $X_{i,T}$,
$X_{i,T+1}$, $X_{i,T+2}$, etc. When $H>0.5$, the process has long-term coherence,
with increases in ${\dot M}_{i,T}$ tending to be followed by more increases, and decreases tending to be followed by more decreases \citep[see][]{kelson2014}.
The effect of long-term correlations, as mentioned earlier (Section \ref{sec:markovian}), is often described as ``memory'' or ``persistence'' in individual galaxy SFHs, but
this interpretation is open to debate.\footnote{The surrounding climate that feeds the stochastic
process also fluctuates on a broad range of timescales \citep[perhaps all timescales;][]{pipiras2000}, and this
produces behavior identical in distribution to physical processes that would locally encode or mimic ``memory'' in individual systems \citep{koutsoyiannis2002}.}

Regardless, for a given value of $H$, the mean stellar mass growth rate for the distribution of galaxies in ensemble $I$ is
\begin{equation}
	{\rm E\/}[{\dot M}_{I,T}] = {1\over \sqrt{2\pi}} {t^{H}\over H}{\overline\sigma_{I,T}}
\label{eq:SFR}
\end{equation}
and the expectation value for stellar mass is simply the integral of Equation \ref{eq:SFR}:
\begin{equation}
	{\rm E\/}[M_{I,T}] = {1\over \sqrt{2\pi}} {t^{H+1}\over H(H+1)}{\overline\sigma_{I,T}} + M_{I,0},
\label{eq:mass}
\end{equation}
where, assuming we begin our calculations sufficiently early (Section \ref{sec:sigbar}), the constant of integration $M_{0}$ can be safely set to zero.

Irrespective of the phenomenology of covariance between stochastic changes to galaxy states, Equations \ref{eq:SFR} and \ref{eq:mass} are key
outcomes of the underlying nature of galaxy growth occurring in states of quasi-static equilibrium with correlated fluctuations.

The square root of the second moment of $P({\dot M}_{I,T})$ can also be calculated \citep[e.g.][]{mandelbrot1968}:
\begin{equation}
	{\rm Sig\/}[{\dot M}_{I,T}] = H^{1/2} {\rm E\/}[{\dot M}_{I,T}]
\label{eq:SFRscat}
\end{equation}

Together Equations \ref{eq:SFR}, \ref{eq:mass}, \ref{eq:SFRscat} combine to yield a star forming main sequence \citep{noeske2007} defined by
\begin{equation}
	\langle{\rm sSFR}\rangle \propto {\rm E\/}[{\dot M}_{I,T}/M_{I,T}] = \frac{H+1}{T}
\label{eq:SSFR}
\end{equation}
with a scatter of
\begin{equation}
	{\rm Sig\/}[{\dot M}_{I,T}/M_{I,T}] = H^{1/2} {\rm E\/}[{\dot M}_{I,T}/M_{I,T}]
\label{eq:SSFRscat}
\end{equation}

These are general forms that show that distributions of galaxies in specific star formation rate are
independent of ${\overline\sigma}_{I,T}$, independent of the galaxy-level $\sigma_{I,T}$ stochastic fluctuations, even independent of $I$,
i.e., all the ensembles overlap each other in their distributions of ${\dot M}_{I,T}/M_{I,T}$.
Hence, Equations \ref{eq:SSFR} and \ref{eq:SSFRscat} amount to functional predictions for the zeropoint and scatter of the SF main
sequence that not only do not require, but are in fact {\it independent} of astrophysics. We return to this point in Section \ref{sec:disc}, below.


\subsection{The Hurst Parameter}
\label{sec:hurst}

Equations \ref{eq:SSFR} and \ref{eq:SSFRscat} have have been compared to available data in \citet{kelson2014}, who found
that published data from the literature on the star forming main sequence constrained the Hurst parameter to be $H=0.98\pm 0.06$.
With $H=1$, and correcting for mass loss using a \citet{salpeter1955} IMF,
one obtains ${\rm Mean\/}[{\rm SSFR\/}]\approx 4/t$, what \citet{madau2014} also inferred for galaxies in the high redshift universe
from the evolution of the star formation rate and stellar mass densities. The derivations for the second moment also yield a
uniform intrinsic scatter of $0.3-0.4$ dex in SSFR, which clearly propagates into the scatter of expected mass growth.
Accurate measurement of this scatter will be sensitive to potential sample selection biases and to measurement methodologies, but this
prediction is also consistent with the data \cite[e.g.][]{salim2007,speagle2014,schreiber2015}. For the remainder of the paper we adopt $H\equiv 1$.

\subsection{Linearly Rising Star Formation Histories at Early Times}

We wrote down a set of equations describing star formation histories as a stationary stochastic process, but it is important to note that a stationary process is not a static process.
The SFRs of individual objects may wander far and wide while, for each ensemble $I$, the mean (and sum) of these {\it changes\/} at fixed $T$
--- $\langle X_{i\in I}\rangle$, $\sum_{i \in I} X_{i}$ ---
are identically zero (Equations \ref{eq:stmean} and \ref{eq:stationary}).
This identity, again, is consistent with notions of steady state equilibrium but also with conservation of mass, even in unsteady states,
when the timescales to convert baryons to stars are inconsistent with the timescales for stochastic changes to galaxy states \citep[][see Appendix \ref{sec:app4}]{little2008}.

Over time, as individual SFRs, $\dot{M_{i}}$, migrate away from their initial starting conditions of ${\dot M}_{i,0}=0$ via Equation \ref{eq:three},
the mean of the ensemble ($\langle \dot{M_{I}}\rangle$), simply increases linearly with time via Equation \ref{eq:SFR}. Put more simply,
anybody analyzing representative {\it samples} of galaxies would then infer linearly rising mean star formation histories:
\begin{equation}
	{\rm E\/}[{\dot M}_{I,T}] = { {\overline\sigma_{I,T}}\over \sqrt{2\pi}} T
\label{eq:SFR1}
\end{equation}

It may seem like sleight-of-hand to start with star formation rates of zero, assert that the mean change in star formation rates is zero
(Equation \ref{eq:stationary}), and arrive at mean SFR histories that are themselves nonzero.
Yet, this is a natural consequence of the fact that, by definition, SFRs cannot be negative, while SFR {\it changes\/} obviously can.
By only allowing SFRs to fall to zero (and no lower) --- by asserting nonnegativity (${\dot M}\ge 0$) in Section \ref{sec:meanevol} ---
we impose a physical prior that bounds the evolution of the distribution.
If we had {\it not\/} done this, the expectation value of the stochastic process simply stays constant,
with ${\rm E}[{\dot M}]\equiv 0$ (such as in classical Brownian motion), and an expectation for the scatter that grows as ${\rm Sig}[{\dot M}] \propto t^{H}$.
With nonnegativity, the first moment of the probability distribution naturally tracks the growth of the second moment of the probability distribution,
moving to higher values as the distribution of $\dot{M}$s broadens away from zero over time.
As a result, nonnegativity is the reason the expectation values grow with time away from zero.\footnote{It is also the reason why the relative scatter
stays constant with time, at ${\rm Sig}[{\dot M}]/{\rm E}[{\dot M}]=\sqrt{H}$.}

Thus when fuel for growth is abundant, the average star formation rate for an unbiased sample of galaxies ---${\rm E}[\dot{M_{I}}]$ ---
will grow linearly in time since the beginning of star formation. At the earliest epochs there is undoubtedly a
dispersion in start times, but the superposition of all such ensembles of galaxies should largely appear to have linearly growing SFRs
given the insensitivity of Equation \ref{eq:SSFR} to how $I$ is chosen.
This is not to say that the SFRs of individuals grow linearly, but, if they could be tracked, we would see that the mean SFR of
the full sample rises in this manner (and so how one might infer this if individual SFHs are based on bulk observations).
One can already see how this behavior might connect to the early rise of the cosmic star formation
rate density \cite[e.g.,][and Section \ref{sec:mfs}, below]{madau2014}.

As a corollary, these results imply that the mean stellar mass of that sample of early galaxies --- simply the integral of Equation
\ref{eq:SFR1} --- must rise {\it quadratically\/} in time:
\begin{equation}
	{\rm E\/}[M_{I,T}] = {{\overline\sigma_{i,T}}\over \sqrt{2\pi}} {T^2\over 2}
\label{eq:mass1}
\end{equation}
This is somewhat remarkable: Though, again, some individual galaxies will
grow faster or slower than this, the {\it galaxy population} matures quite rapidly, no doubt bringing with it the usual byproducts of stellar mass
growth --- metals, dust, energy from supernovae --- all to be deposited into the CGM and IGM.


\subsection{Next Steps: Towards the Galaxy Stellar Mass Function}

At early times no significant dependence of SSFRs on galaxy mass has been observed \cite[e.g.][]{gonzalez2014},
and there is abundant fuel to drive star formation globally. Therefore
this mathematical description ought to fully describe the ensembles of galaxies at those epochs. Unfortunately, any predictions for the
distributions of masses and star formation rates are normalized by whatever the expected distributions of
$\overline\sigma$ might be (e.g., Equation \ref{eq:SFR}). These RMS stochastic fluctuations define the absolute mass scales for ensembles of galaxies and
their constituents.
While these stochastic fluctuations cancel-out when dividing ${\dot M}$ by $M$, e.g., for generating the star forming
main sequence, one really, {\it really\/} would like to be able to
derive them from first principles, too. That exercise is the focus of the rest of the paper.


\section{The Distribution $P({\overline\sigma})$ and the Emergence of the Stellar Mass Function}
\label{sec:sigbar}


In the previous section we concluded that ensembles of galaxies growing stochastically will have mean rates
of star formation and stellar mass that scale with time in simple, predictable ways, with the caveat that
the underlying distributions of growth histories as of yet have unknown absolute normalizations.
In this section we relate the normalizations of the mean growth histories to analytical approximations for the second derivative of baryon accretion
onto dark matter halos at the onset of star-formation, i.e., the conditions of the halos when atomic cooling can commence
\citep[e.g][]{silk1977} and modern stars can be formed.\footnote{Population III stars can form in lower-mass halos where molecular --- not atomic---
hydrogen is the dominant coolant \citep{yoshida2003}. Though technically this phase represents the ``instantiation'' of star formation,
we neglect it here because it is by definition distinct from the mode that characterizes the long-term growth and maturation of galaxies (which we seek to describe).
As soon as the metals from the first few generations of Pop III stars pollute the ISM, sustained Pop II/I star formation takes over, permanently.
Hence, the omission of such low-mass halos (and Pop III production generally) should not invalidate the analytical arguments that follow,
which may be taken to describe Pop II/I stellar mass production.}

At root, we seek to derive (1) a characteristic value for the RMS stochastic fluctuation amplitudes, which we will
call ${\overline \sigma}^*$; (2) a power-law distribution, $P({\overline \sigma})$, around ${\overline \sigma}^*$;
and (3) the normalization, or amplitude of $P({\overline \sigma})$. Because we derive these sequentially, and attempt to
cover a lot of ground, we note the key implications of this section here and encourage readers to keep them in mind as they
progress:

\begin{itemize}
\item The stellar mass function is related to the distribution of ${\overline\sigma}$ through the
equivalence $P({\overline\sigma})\equiv \phi(2\sqrt{2\pi}M/t^2)$;
\item The physical values of ${\overline\sigma}$ control the mean rate of stellar mass growth;
\item The shape and normalization of $P({\overline\sigma})$ is more sensitive to the spectrum of small scale {\it density} fluctuations
than to the distribution of halo {\it masses} beginning to form stars;
\item While the power-law slope to low ${\overline\sigma}$, $\alpha$, depends on the
anisotropy of the accretion, in all cases $\alpha>-2$, with precisely $\alpha=-7/5$ for isothermal spheres
accreting material through filaments whose diameters are not correlated with the masses of the halos they are feeding;
\item No excess feedback mechanisms appear to be required to make stellar mass functions shallower than the dark
matter halo mass function.
\end{itemize}

Immediately following our derivations of ${\overline \sigma}^*$, $\alpha$, and the overall normalization,
we can then use the probability distribution $P({\overline\sigma})$ to
calculate expectations for the distribution and evolution of star formation rates and stellar masses at early times (Section \ref{sec:mfs}).

\subsection{Relating $\overline\sigma$ to the Acceleration of Stellar Mass Growth}

Let us first rewrite Equation \ref{eq:SFR1}, now explicitly switching to notation of ${\overline\sigma}_{I,t}$ --- describing galaxy ensemble $I$ at time $t$ ---
for the remainder of the presentation:
\begin{equation}
	{\rm E}\left[\dot{M}_{I,t}\right]={\rm{E}}\left[ \frac{dM}{dt}\right]_I =\frac {\overline{\sigma}_{I,t}}{\sqrt {2\pi }} t.
\label{eq:Macc}
\end{equation}
Specifically, the mean SFR of ensemble $I$ is expected to increase linearly with time, so long as ${\overline\sigma}_{I,t}$ is itself
not a function of time. Thus one can think of ${\overline\sigma}_{I,t}$ as a sort of mean acceleration\footnote{Where
Equation \ref{eq:Macc} is analogous to ``$v=a\times t$''}, foreshadowing what is to come.

We write the derivative of the mean SFR with respect to time:
\begin{equation}
\frac {d}{dt}{\rm E}\left[ \frac{dM}{dt}\right]_I =
	\frac{\overline{\sigma}_{I,t}}{\sqrt {2\pi }} + \frac {t }{\sqrt {2\pi }}\frac{d {\overline\sigma}_{I,t}}{d t},
\end{equation}
or:
\begin{equation}
{\rm E\/}\left[ \frac{d^2M}{dt^2}\right]_I =
	\frac{\overline{\sigma}_{I,t}}{\sqrt {2\pi }}\left(1 + t \frac{d\ln {\overline\sigma}_{I,t}}{dt} \right),
\label{eq:d2mdt2}
\end{equation}
because integration and differentiation are linear operators, allowing us to move the differentiation inside the integration for the expectation value.
(Recall: though the individual SFHs are not differentiable, the expectation values are continuous and smooth, and therefore {\it are\/} differentiable; Section \ref{sec:stochDef}.)
Equation \ref{eq:d2mdt2} represents a key result.

\subsection{Simple Ensemble Evolution}

In this paper we are only going to investigate the behavior of systems in which ${\overline\sigma}_{I,t}$ is constant with time.
In essence, this means examining epochs where no external phenomena --- such as the availability or distribution of fuel supplies --- restrict $X_{i,t}$,
or how galaxies explore SFR space. In this case, the second term in Equation \ref{eq:d2mdt2} is assumed to be identically zero.
If our interpretation is correct (and we will empirically show evidence for this) such an assumption is valid at early cosmic times.
However, this does not diminish the importance of the second term, about which we speculate in Section \ref{sec:late} and Appendix C, below.

In this case:
\begin{equation}
        {\rm E}\left[ \frac {d^{2}M}{dt^{2}}\right]_I = \frac {{\overline\sigma}_{I,t} }{\sqrt {2\pi }}.
\label{eq:sigeq}
\end{equation}

Equations \ref{eq:d2mdt2} and \ref{eq:sigeq} allows us to rethink $\overline{\sigma}$ not as an hypothetical RMS fluctuation, but as a quantity related to the time
derivative of the average mass flux into galaxies (or their halos). Because of our assumption of time-invariance for ${\overline\sigma}_{I,t}$, we take Equation
\ref{eq:sigeq} to be valid {\it at the start of star-formation\/} in particular.

Given a set of initial conditions ${\dot M}=0$ and $M=0$ at $z_{\rm start}$, we can derive the unique mapping of halo conditions for ensemble $I$ to
${\overline\sigma}_{I,z_{\rm start}}$\footnote{By definition ${\dot M}=0$ and $M=0$ prior to $z_{\rm start}$, as the conditions to support the sustained formation of stars have
not yet been generated}. These estimates of ${\overline\sigma}$ are explicitly tied to the dark matter accretion for a given halo mass
at $z_{\rm start}$, modulo a star-formation efficiency and the baryon fraction.

It is both particularly interesting and fortuitous that we do not require knowledge about the subsequent accretion histories of the halos that instantiate star formation.
It is fortuitous because all dark matter halos of a given mass at $z_{\rm start}$ do not comprise a representative ensemble of halos obeying analogous (dark matter) growth
equations to those derived in Section \ref{sec:stoch}. They would need to be complete (or at least representative of) ensembles that arose out of similar initial conditions
so that the expectation values we derived would remain valid. Halos that will begin forming stars at $z_{\rm start}$ began their $H=1$ stationary stochastic growth
significantly earlier, when a range of wavenumbers and mass scales went nonlinear (\citealt[][in prep]{kelson2017}). As such, halos of a given mass at $z_{\rm start}$ are drawn
from a diversity of ensembles each with mean ensemble growth trajectories specified by initial conditions set in place at those earlier epochs. In principle, one could
begin constructing halo mass functions out of a similar framework to that presented in this paper, beginning growth at much earlier times, but these efforts are
deferred until a later date.

Note that though the ensembles evolve deterministically, the member galaxies do not, as dark matter (and baryon) accretion proceed stochastically
\citep[e.g.][]{bond1991,kelson2017}. The continued stochastic dark matter/baryon accretion is, of course, (1) supplying the fuel for stellar mass growth, (2) imposing the
events that stochastically change stellar mass growth rates, and (3) correlating those stochastic changes through the power spectrum of density inhomogeneities.
Put more simply: in a big enough volume, enough modes are present such that the {\it distribution\/} of future accretion histories is already written.

As stated, this approximation is valid when the universe can support unabated fueling of galaxy growth. The quadratic growth in mean mass ensures that the baryon content
of the universe is depleted roughly linearly with time (ignoring recycling), so one expects the diminishment of available baryonic material to eventually degrade the
long-term expectations of the stochastic process. In other words stationarity will, {\it at some point\/}, be violated. Such issues are reserved for Section \ref{sec:late}.


\subsection{Approximating $\overline\sigma$ with Baryon Accretion Rates At the Start of Star Formation}

Let us assume that at $z\sim 10$--15 we can write that the rate of formation of stellar mass is proportional to the inflow of baryonic matter, such that $dM/dt=\epsilon\times dM_b/dt$.
Then we can write that the change in stellar mass is proportional to the physical baryon density, $\rho_{b}$, times the average infall velocity of the accreting baryons,
$v_{b}$, onto whatever dark matter halos begin forming stars.

We will assume there is one galaxy per halo of mass $M_{h}$, but we will {\it not} assume a 1:1 stellar-to-halo-mass connection.
Instead, we will derive the {\it distribution} of stellar masses for a given $M_{h}$ and show that---when convolved with a much broader kernel reflecting the impact of the
local over-density distribution---this gives rise to the stellar mass function and its evolution.

We start with the continuity equation
\begin{equation}
	\dfrac {d\rho_b}{dt}=-\frac{1}{r^2}\frac{\partial}{\partial r}{\left(r^2\rho_b v_b\right)}.
\end{equation}
and take the first derivative,
\begin{equation}
	\dfrac {d^2\rho_b}{dt^2}=-\frac{1}{r^2}\frac{\partial}{\partial r}{\left(r^2\rho_b \frac{\partial v_b}{\partial t}\right)}.
\end{equation}

Outside of the halo's radius, $r_h$, we characterize the density distribution of baryons as a power-law for simplicity \cite[see e.g.,][]{bertschinger1985,shapiro1999}:
\begin{equation}
	\rho_b=f_b \rho_{h}\left(\frac{r}{r_h}\right)^{-\gamma},
\end{equation}
where $\gamma>0$, $f_b$ is the cosmic baryon fraction, and $\rho_{h}$ is the mean density of the halo within $r_h$.
Applying Newtonian gravity---$\partial v_b/\partial t\propto GM_h/r_h^2$---and substituting, we have:
\begin{equation}
	\dfrac {d^2\rho_b}{dt^2}= \frac{\gamma f_b GM_h\rho_{h}}{r_h^3}\left(\frac{r}{r_h}\right)^{\gamma-3}.
\label{eq:d2rhodt2}
\end{equation}

Using this form, we can express the second derivative of the baryonic mass of a halo experiencing three types of accretion at the instantiation of star formation:
\begin{enumerate}
	\item[] {\bf Case 1}: isotropic accretion;
	\item[] {\bf Case 2}: accretion onto individual halos through surface areas proportional to individual halo surface area;
	\item[] {\bf Case 3}: accretion onto individual halos through a mean surface area that is uncorrelated with halo mass or local density.
\end{enumerate}
The latter two could be characterized as equivalent to accretion along streams or flows \cite[e.g.][]{dekel2009}.
In Case 2, these flows have diameters that depend on the halo they are feeding, in Case 3 they do not.

Case 3 thus applies when the growth of massive halos has decoupled from the growth of the halos that occupy filamentary
tributaries. Such accretion flows can be said to have a fixed metric area, or fixed diameter, as opposed to fixed covering fractions.  
Note: the term ``fixed'' here {\it does not imply constant with time\/} but {\it constant with respect to halo properties\/}.
We take Case 3 as our default accretion mode.

For the growth of stellar mass in halos that are accreting isotropically, the second derivative of the baryonic mass of the halo has the form:
\begin{equation}
	\dfrac {d^2M_b}{dt^2} = \frac{4\pi r_h^2 l_{\rm ff} \gamma f_b GM_h\rho_{h}}{r_h^3},
\label{eq:iso0}
\end{equation}
where $l_{\rm ff}$ is distance the accreted baryons traverse in a free-fall time. Because $l_{\rm ff}\equiv r_h$, and
because the halo may be permeable to accretion only through a fraction, $C$, of the total surface area, one obtains:
\begin{equation}
	\dfrac {d^2M_b}{dt^2} = 4\pi C \gamma f_b GM_h\rho_{h}.
\label{eq:iso}
\end{equation}
When $C=1$ the accretion is isotropic (Case 1), else it can be thought of as being due to inflowing streams whose areas
are proportional to the surface area of the halo (Case 2, above).

For a fixed metric diameter, or cross section of accretion (Case 3), Equation \ref{eq:d2rhodt2} becomes:
\begin{equation}
	\dfrac {d^2M_b}{dt^2}= \frac{N_s \pi r_s^2 l_{\rm ff} \gamma f_b GM_h\rho_{h}}{r_h^3}.
\label{eq:metric0}
\end{equation}
Here $N_s$ denotes the number of accretion streams with mean cross section $\pi r_s^2$.
If the total cross sectional area of these accretion streams equals the surface area of the spherical
halo, the result is equivalent to the isotropic case. When the width of the typical accretion stream
is independent of halo mass, such as when the flow is defined by local radiative and hydrodynamical
instabilities, one obtains:
\begin{equation}
	\frac{d^2M_b}{dt^2}= \frac{N_s \pi r_s^2 \gamma f_b GM_h\rho_{h}}{r_h^2}.
\label{eq:metric}
\end{equation}

Since $r_h^3=3M_h/(4\pi\rho_h)$, the above becomes:
\begin{equation}
	\frac{d^2M_b}{dt^2}= \left(\frac{4\pi}{3}\right)^{2/3} N_s \pi r_s^2 \gamma f_b G M_h^{1/3}\rho_{h}^{5/3}.
\label{eq:metric2}
\end{equation}
These halos have only just collapsed, in a \citet{press1974} sense, and therefore have a mean
matter density $\rho_{h}=178\times\Omega_M(z)\rho_c $.

Thus, the second derivative of the stellar mass, as in Equation \ref{eq:sigeq}, for Case 3 is:
\begin{equation}
	\frac{d^2M}{dt^2}= \left(\frac{4\pi}{3}\right)^{2/3} N_s \pi r_s^2 \gamma \epsilon_{*} f_b G M_h^{1/3}\rho_{h}^{5/3}
= \frac{\overline\sigma}{\sqrt{2\pi}},
\label{eq:metrics}
\end{equation}
where $\epsilon_{*}\approx 24\times \epsilon_{\rm ff}$ \citep{dekel2013a}.

Equation \ref{eq:metrics} thus explicitly relates ${\overline\sigma}$ to the halo masses and their ambient matter densities
{\it at the time when star formation begins\/} for a given ensemble of objects. We call this epoch $z_{\rm start}$ but
it is misleading to think of this as a single moment when all halos begin forming stars. Halos will reach a particular mass scale, say one
where the virial temperature is sufficient for atomic cooling to begin, at different times depending on whether they formed in a high
density region of the universe or a low density one. In principle we ought to model how $z_{\rm start}$ depends on large-scale over- or under-density but 
we are trying to keep this paper's basic illustration of the origin and stochastic growth of the stellar mass function as simple as possible. Therefore, for simplicity, we
adopt a single $z_{\rm start}$. The real cosmic distribution of galaxies will be a superposition of many stochastic ensembles, each with intrinsic distributions
that should, in scheme, be similar to what we have derived.

In any event, Equations \ref{eq:sigeq} and \ref{eq:metrics} show that
when baryon accretion occurs along streams whose average cross-sections are independent of halo mass:
\begin{multline}
	{\overline\sigma}_{I,z_{\rm start}} =
\sqrt{2\pi} \left(\frac{4\pi}{3}\right)^{2/3}N_s \pi r_s^2 \gamma \epsilon_{*} f_b  \\ \times\,G M_{h,I,z_{\rm start}}^{1/3}\rho_{h,I,z_{\rm start}}^{5/3}.
\label{eq:sigdep}
\end{multline}
The reader should note the dependencies of ${\overline\sigma}$ on halo conditions
at $z_{\rm start}$. From this one infers that $P({\overline\sigma})$ should likely be only mildly sensitive to the
distribution of halo masses beginning star formation, but significantly more sensitive to the distribution of local
density fluctuations providing fuel. These dependencies are explored at depth in Section \ref{sec:alpha}.

In contrast, halos accreting through fixed fractions of their surface areas would experience stellar mass growth according to:
\begin{equation}
	{\overline\sigma}_{I,z_{\rm start}} =
2\left({2\pi}\right)^{3/2}
C \gamma \epsilon_{*} f_b GM_{h,I,z_{\rm start}}\rho_{h,I,z_{\rm start}}.
\label{eq:isosigdep}
\end{equation}
These two cases result in different dependencies on halo mass and local matter density. Predictably, propagating the halo mass function and power spectrum of density fluctuations through these
two cases will lead to different expectations for the stellar mass function, explored below. 

Note that not all galaxies in ensemble $I$ may have the same halo mass $M_{h,I,z_{\rm start}}$ at $z=z_{\rm start}$, or be embedded in
similar fields of local physical matter density $\rho_{I,z_{\rm start}}$. We expect, however, that either
Equation \ref{eq:sigdep} or \ref{eq:isosigdep}
will allow us to take those galaxies in ensemble $I$ that belong to halos of mass $M_h$ at $z=z_{\rm start}$, and to propagate
the earlier expressions for how their distributions of star formation rates and stellar masses will evolve with time.

Given a spectrum of halo masses, the mean universal density, and a power spectrum of density fluctuations,
we can derive the three parameters that define the
initial ``${\overline \sigma}$ function''---and thus the galaxy stellar mass function of an ensemble and its evolution with time:
${\overline\sigma}^*$, $\alpha_ {\overline\sigma}$, $\int d{\overline\sigma} P({\overline\sigma})$.


\subsection{Deriving $M^*$: A Characteristic Value for $P({\overline\sigma})$}
\label{sec:charsig}

We showed above that ${\overline\sigma}$ is a function of halo mass and the physical density of the surrounding material.
Thus we expect a probability distribution of $P({\overline\sigma})$ with a shape guided by the distributions of both halo
mass and the spectrum of mildly nonlinear density fluctuations around those halos. The halo mass function has a knee, similar
to \citet{schechter1976} functions, and a power-law slope of $\alpha=-2$ to low masses. We also expect a power-law of
densities with a cutoff above some density fluctuation scale defined by the comoving radius that enclosed
a mass equivalent to a given halo mass (see below).

Equation \ref{eq:sigdep} implies a characteristic value of $P({\overline\sigma})$ --- setting the scale of stellar masses --- for a given halo mass:
\begin{eqnarray}
{\overline\sigma}^* =
\left(\frac{\epsilon_{\rm ff}}{0.015}\right)
\left(\frac{\gamma}{2}\right)
\left(\frac{f_b}{0.16}\right)
\left(\frac{N_s}{3}\right)\left(\frac{r_s}{1 \text{\ kpc}}\right)^2
\times\qquad\qquad\cr
\left(\frac{1+z_{\rm start}}{1+12.5}\right)^5
\left(\frac{M^*_{h,z_{\rm start}}}{1\times 10^9\, \Msun}\right)^{1/3}
\left(2.9 \times 10^{-7}\, \Msun\,\text{yr}^{-2}\right).\qquad
\label{eq:sigstar}
\end{eqnarray}
At a redshift for the onset of star formation of $z=12.5$, the
characteristic halo mass is $M^*_{h,z=12.5}\sim 1\times 10^9\, \Msun$ \citep[e.g.,][]{warren2006,tinker2008}.
Here we have also assumed 3 accretion streams \cite[e.g.,][]{dekel2009} with (circular) cross section of radius 1 kpc,
and $\epsilon_{\rm ff}=0.015$ for the star-formation efficiency per free-fall time \citep{krumholz2005}.
Note, too, the adoption of $\gamma=2$ for these earliest dark matter halos, prior to the initiation of star formation
\cite[][]{bertschinger1985,shapiro1999}.

It is interesting to note that for $T=10^4\, \rm K$ gas at that redshift and at the mean density of collapsed halos,
the Jeans length is $\sim 1.2$ kpc, of order the characteristic radius of the adopted fueling streams.
The characteristic halo mass of $10^9\, \Msun$ has a spherical radius of $r_h\approx 1.6$ kpc,
but the halo mass at the cooling limit is $\sim 10^7\, \Msun$ has $r_h\approx 0.34$ kpc. Thus
the total surface area of the streams, $N_s\pi r_s^2$, is larger than the surface area of the lowest mass
halo potentially able to cool.

We take this tension as annoying but not necessarily fatal, given that our derivations have aspects that may only be good, perhaps, to a factor of two.
Moreover, as is clear above, there is degeneracy between $\epsilon_{*}$, $\gamma$, and $N_s \pi r_s^2$. Fortunately for us, and the reader,
detailed explorations of the covariance between those variables is beyond the scope of this paper.
However, it may be that for halo masses with radii comparable to a Bonner-Ebert or Jeans radius, the accretion is
isotropic, and above a particular halo mass scale the accretion is confined to streams defined and set up when those halos were smaller,
such that the streams do not grow in width as the halos grow. Thus there may be a (low halo mass) regime with
dark matter halos experiencing accretion described by Equation \ref{eq:isosigdep}, with a transition to a (high halo mass) regime
with accretion dependent on mass and density described by Equation \ref{eq:sigdep}.
Appendix \ref{sec:app1} derives ${\overline\sigma}^*$ assuming isotropic accretion (Case 1) or accretion through fixed fractions of halo surface area (Case 2).


\subsubsection{Implications for Mean Stellar Mass Growth}
\label{sec:mgrowth}

Using the characteristic value ${\overline\sigma}^*$ from Equation \ref{eq:sigstar} with the
evolution of the mean mass for the ensemble in Equation \ref{eq:mass1}, and assuming a \cite{salpeter1955} IMF,
we now have the time evolution of the characteristic stellar masses of galaxies:
\begin{eqnarray}
	M^*(t) \sim 4 \times 10^{10} \Msun \times \left(\frac{t}{\text{Gyr}}\right)^2.
\label{eq:mstar}
\end{eqnarray}

We cannot stress the importance of this result enough. Given only the basic assumptions of (1) stationary stochastic star
formation with correlated fluctuations, and (2) the Newtonian gravitational attraction of baryons by early dark matter halos,
we have derived the evolution of the characteristic stellar mass of galaxy ensembles --- the mass function's $M^{*}$ --- to arbitrary look-forward times
(while fuel supplies are ample). This must be a core aim of any basic theory of galaxy evolution, yet it has been achieved {\it with notable ignorance of any astrophysical details\/}.
Furthermore, we have at no point assumed that any individual galaxies in the ensemble obey this mean mass growth. Underpinning the growth of the ensemble is an infinite
set of quite diverse histories satisfying the definition of a stationary stochastic process.

Using Equation \ref{eq:mstar}, one finds that ensembles of galaxies at times 0.1 Gyr, 0.5 Gyr and 1.5 Gyr after the instantiation of star formation
($z\sim10.5$, 6.5, and 3.5, \resp) should have characteristic stellar masses $M^*\sim 4\times 10^{8}\, \Msun$, $1\times 10^{10}\, \Msun$, and $9\times 10^{10}\, \Msun$,
consistent with data where they exist (Section \ref{sec:mfs}, Figure \ref{fig:mfevol}). This rapid increase in the average stellar mass content of the universe
should also be reflected by a rapid pollution of galaxy environments with metals and dust.


\subsubsection{Implications for Dependence on Environment}

Of course, any observed stellar mass functions will be superpositions of (1) ensembles originating from a range of
${\overline\sigma}^*$s from different large-scale over- and under-densities, as well as (2) ensembles
whose halos begin forming stars at different times because they do not reach appropriate conditions
for cooling simultaneously.

Each region of the universe will sustain its own $\sigma^*$, and thus $M^*$, through the term $\rho^{5/3}$. In other words,
the mean mass growth of an ensemble of galaxies will depend quite strongly on the macro environments of the galaxy seeds.
Through this term, the masses and SFRs of galaxies will grow more slowly in under-dense
regions of the environment than in over-dense regions of the universe. To illustrate this point, let us rewrite
Equation \ref{eq:sigdep} as: 
\begin{multline}
	{\overline\sigma}_{I,z_{\rm start}} =
\sqrt{2\pi} \left(\frac{4\pi}{3}\right)^{2/3} N_s \pi r_s^2 \gamma \epsilon_{*} f_b G M_{h,I,z_{\rm start}}^{1/3} \times\\
\left[178\Omega_M(z_{\rm start})\rho_c(1+\delta_{\rm LS})(1+\delta_{\rm SS})\right]^{5/3},
\label{eq:deltasigdep}
\end{multline}
explicitly invoking large scale ($\delta_{\rm LS}$) and small scale ($\delta_{\rm SS}$) density fluctuations. Here large scale 
is defined as $R\gg \left(3M_h/4\pi \rho_{c}\right)^{1/3}$.

More concretely, in regions of the universe over-dense by a factor of $\sim 1.5\times$ ($\delta_{\rm LS}=0.5$), $P({\overline\sigma})$ would have a characteristic
${\overline\sigma}^*$ higher by a factor of $1.5^{5/3}\simeq2$ than the global mean.
If ensembles of halos in such high density regions initiated star formation at the same
time as average regions of the universe, then one would see the characteristic masses of halos in the two regions grow according to
$M^*(t) \sim 4 \times 10^{10} (1+\delta_{\rm LS})^{5/3}\, \Msun \times \left(t/\text{Gyr}\right)^2$.
This enhancement in growth rates and masses --- i.e., assembly bias \citep[e.g.][]{wechsler2006},
or the accelerated evolution of over- as compared to under-dense regions \citep[e.g,][]{dressler1980, abramson2016}
--- would hold presumably until a point when the universe could no longer sustain unabated growth in such a large scale over-density, due, perhaps, to diminished cold gas supplies.

At such a time, additional physics would alter the long-term expectations of star-formation through the second term of Equation
\ref{eq:d2mdt2}. While we speculate further about its meaning in Section \ref{sec:mfs} and Appendix C,
the form of these additional terms is beyond the scope of this work. Note however that the dependence of ${\overline\sigma}^*$ on $(1+\delta_{\rm LS})^{5/3}$
should follow straight through to global trends of galaxy properties with environment such as the morphology-density
relation \citep{dressler1980}, or the dependence of clustering on galaxy types and its evolution \citep{coil2008}, so long as $(1+\delta_{\rm LS})$ correlates
with with large-scale environment --- i.e. scales significantly larger than the turnaround radii of the halos instantiating star formation\footnote{Note that on smaller scales,
there will be scatter introduced by the nonlinear growth of structure.}


\subsection{Deriving $\alpha$: A Power-Law Slope for $P({\overline\sigma})$}
\label{sec:alpha}

In the estimates given above for the characteristic stellar masses of galaxies, we only needed the characteristic halo masses
at $z_{\rm start}$ for the seeds of galaxies. In deriving theoretical stellar mass functions below, we will use the
full halo mass functions at $z\sim 10-15$ as the starting point, scaling ${\overline\sigma}^*$ appropriately for each halo
mass above the atomic cooling limit \citep[e.g.][]{silk1977}. Deriving the expected distribution of stellar masses that may arise for
a given seed halo mass also requires estimating the power-law slope to low ${\overline\sigma}$, as well as an overall normalization.
We now derive these properties of $P({\overline\sigma})$.


\subsubsection{The Sub-dominance of the Dispersion in Halo Masses}
\label{ref:mhalo}

We have derived a dependence on $M_{h,z_{\rm start}}$ that is quite mild, with ${\overline\sigma} \propto M_{h,z_{\rm start}}^{1/3}$. Given that
the halo mass function is very steep, $\phi(M)\propto M^{-2}$, one might have expected something like $P({\overline\sigma})\sim
{\overline\sigma}^{-4}$ from propagating the halo mass function through the chain rule. However, at the redshifts where star formation is
initiated, $z\sim 10$, the range of halo masses with virial temperatures that can support atomic cooling
\citep[$T_{vir}>10^4$K; e.g.][]{rees1986,efstathiou1992} spans {\it
only\/} 2-3 dex, from the cooling limit ($\sim 10^7 \Msun$) up to the characteristic halo mass. And even with the uncertainties on this floor,
the exponent of $1/3$ reduces the range of star forming halo masses to a span in $\overline\sigma$ that is significantly smaller, $\simlt 1$ dex.

In other words, when deriving the full stellar mass function of galaxies in this formalism, the top end of the halo mass function only
acts like a narrow kernel, convolving the much broader dependence of $P({\overline\sigma})$ that, as we will show below, comes from the distribution of (local) matter densities.

\subsubsection{The Dominance of Local Matter Density}

From Equations \ref{eq:sigdep} and \ref{eq:isosigdep}, we see that the dependence of $\overline\sigma$ on the density of the matter being accreted is
\begin{equation}
{\overline\sigma} \propto \rho^{x/3}
\label{eq:x3}
\end{equation}
where $x=5$ for a distribution of halos with accretion flows whose mean diameter is mass-independent (Case 3),
and $x=3$ for accretion through a fixed fraction of halo surface area (Cases 1 and 2; see Section \ref{sec:sigbar}). In the former,
${\overline\sigma}\propto M_{h,z_{\rm start}}^{1/3}$. As the distribution
of halo masses is not very sensitive to $\delta_{\rm LS}$ at these epochs \citep[e.g.,][]{mo1996}, the rate of accretion onto
those halos is very much more sensitive to $\delta_{\rm SS}$, the highly local matter density, than $M_{h}$ ---  {\it five times more sensitive\/}, in the log.

In the $x=3$ case (Equation \ref{eq:isosigdep}),
the sensitivity  of ${\overline\sigma}$ to halo mass and local matter density are similar, though,
as will be shown below (Section \ref{sec:norm}),
the span of halo masses available for cooling is overwhelmed by the range of matter densities, leaving the power-law
slope of $P({\overline\sigma})$ still dominated by the distribution of density fluctuations, not halo masses.

\subsubsection{The Spectrum of Local Density Fluctuations}

We derived above how the spectrum of local density fluctuations must drive the distribution of ${\overline\sigma}$s, even more than
the distribution of halo masses can. In order to derive the power law distribution of ${\overline\sigma}$,
we must know the probability distribution for lumps of matter of a given scale around each halo about to initiate star formation.
This distribution is described by the two-point correlation function, $\xi(r)$, which can be obtained from the
power spectrum $P(k)\sim k^n$ \citep[e.g][]{peebles1974}. When the power spectrum is a power law of index $n$, the correlation function is:
\begin{equation}
	\xi(r) \propto r^{-(n+3)}.
\label{eq:corr}
\end{equation}
\citep[see][]{peebles1975}.

Ultimately, we wish to use the chain rule to translate the two-point correlation function into a distribution $P({\overline\sigma})$ via
\begin{equation}
	P({\overline\sigma})d{\overline\sigma}\propto r^2 \xi(r)dr.
\end{equation}
To do so, we first recognize that on small scales---scales smaller than
the those that collapsed and formed the halos under question---the power spectrum approaches $n\sim -3$. When $n=-3$, the
RMS mass fluctuations are constant with $r$, allowing us to approximate $\rho\propto r^{-3}$. Together with Equation \ref{eq:x3} we therefore have:
\begin{equation}
	{\overline\sigma}\propto r^{-x},
\label{eq:r5}
\end{equation}
where $x=5$ for Case 3 (Equation \ref{eq:sigdep}), and $x=3$ for Cases 1 and 2 (Equation \ref{eq:isosigdep}).

Propagating the chain rule, we have
\begin{equation}
	P({\overline\sigma})\propto {\overline\sigma}^{(n-x)/x}.
\label{eq:nsig}
\end{equation}
Given Equation \ref{eq:mass}, we know
that an ensemble of halos that begins forming stars will produce stellar mass functions $\phi(2\sqrt{2\pi}M/t^2)
\propto P({\overline\sigma})$.
After star formation begins, the low mass end of the stellar mass function should, according to Equation \ref{eq:nsig}, thus follow a power-law of:
\begin{equation}
	\phi(M)\sim M^{(n-x)/x}=M^{\alpha},
\end{equation}
where $\alpha=(n-x)/x$.

So what value of $n$ is most relevant to our derivation? On the smallest scales, $n=-3$,
implying an asymptotic value of $\alpha=-8/5=-1.6$ for $x=5$, and $\alpha=-2$ for $x=3$.
But at the mass scales of the halos being discussed here, the linear power spectrum has an effective power-law slope of $n_{\rm eff}\approx -2.5$ \citep{murray2013}, leading to
expectations of $\alpha\approx -1.5$ or $\alpha\approx -1.8$ for the cases of $x=5$ and $x=3$, respectively.
With such typical values of $n_{\rm eff}$ (on relevant scales) one ought to {\it never\/} expect stellar mass functions with
power-law slopes equal to the low mass slope of the halo mass function itself ($\alpha=-2$).

\subsubsection{The Effects of Non-linearity}

We have already recast ${\overline\sigma}$ using the acceleration of baryons onto halos at the onset of star formation. As stated above,
these are dependent on the local matter density fluctuations, which, on the
relevant scales, are themselves collapsed (sub)halos. Thus, it is precisely the 1-halo term of the
correlation function that tells us the probability
distribution of density fluctuations being accelerated onto our dark matter halos. In other words,
we are interested in the power spectrum in the (mildly) non-linear regime, near and around those halos recently collapsed, e.g., within the turnaround radii of those massive halos instantiating star formation.

The linear power spectrum does not suffice for these purposes, but
there exists extensive analytical and numerical work on the nonlinear power spectrum in this regime
\cite[e.g.,][]{peacock1996,ma2000}, such that we can make progress.

Changing the nomenclature of their exponent to a $\gamma$ for convenience, \citet{ma2000}
found that when the matter density profiles in these new halos are of the form
\begin{equation}
	\rho(r) \sim r^{-\gamma},
\label{eq:rden}
\end{equation}
the 1-halo term of the correlation function follows:
\begin{equation}
	\xi(r) \sim r^{3-2\gamma}.
\end{equation}

This power-law correlation function is then equivalent to a power spectrum of the form:
\begin{equation}
	P(k) \sim k^{2\gamma-6},
\label{eq:pkgamma}
\end{equation}
	such that:
\begin{equation}
	P({\overline\sigma}) \propto {\overline\sigma}^ {(2\gamma-6-x)/x}.
\label{eq:alphax}
\end{equation}

The outer profiles of low redshift dark matter halos
systematically transition from $\gamma=3$ at $r_{200}$ to $\gamma=1$ at distances connected to the Hubble flow
\citep{diemer2014,wetzel2015,lau2015}. While their
halos do not follow a uniform $\gamma=2$ at such large radii, the {\it mean\/} logarithmic slope in those regions is somewhere
between $2.5\simlt \gamma \simlt 1.5$. Assuming the baryonic density profile traced the dark matter, this range of $\gamma$
implies stellar mass functions with $-1.2 \simgt \alpha \simgt -1.6$ if $x=5$, and
$-1.33 \simgt \alpha \simgt -2$ if $x=3$.
The \citet{diemer2014} halos show a trend towards shallower density profiles at higher redshift, at least to $z=6$,
and while it is unclear what the numerical expectations are for $\gamma$ at $z\sim 10$,
they are almost certainly result in values of $\alpha > -2$.

\subsubsection{Our Best Guess}

If the matter density disconnected from the Hubble flow can be described by $\gamma=2$ (i.e., isothermal spheres)
and accretion occurs through fixed diameters that are uncorrelated with halo mass then:
\begin{equation}
	P({\overline \sigma})\sim  {\overline \sigma\ }^{-7/5}.
\end{equation}
resulting in stellar mass functions with $\alpha=-7/5$, i.e., 
{\it precisely the value observed at effectively all epochs\/}. 

The non-singular isothermal spheres
explored by \citet{shapiro1999} for the equilibrium structure of collapsed dark matter halos have $\gamma\approx 2.1$ at the radius of the
halo, implying $\alpha\approx -1.36$. For the \citet{bertschinger1985} solutions, $\gamma\approx 2.25$, leading to $\alpha\approx -1.30$.
Equation \ref{eq:alphax} clearly shows that $\alpha$'s sensitivity to $\gamma$ is mild, with
$\Delta\alpha=\Delta\gamma/5$.

What we have just derived is surprisingly robust given the magnitude of the uncertainties in $\gamma$.
For a reasonable range of density profiles, $\alpha \approx -7/5$, so for the remainder of this work we adopt $\gamma=2$.
Integrating over the full span of halo masses above the atomic cooling limit will not change this expectation.
The logarithmic span of halo masses is (a) not large, and (b) substantially diluted by the power-law distribution of local matter densities
up to the scale characterized by a given halo mass. In other words, the expected power-law distribution of $\overline{\sigma}$, and by extension
the stellar mass function of galaxies, simply should not have, and could not ever have had, a low mass slope of $\alpha=-2$.

Our derivations suggest that previous work required additional mechanisms to flatten the faint-end slope of the mass function because of the assumption
of isotropic accretion. Assuming isotropy with isothermal spheres, Equation \ref{eq:alphax} yields $\alpha=-5/3$, in line with the
$-1.5\simlt \alpha\simlt -1.7$ estimated by \cite{white1991} for their models.

We now {\it almost\/} have a fully defined distribution $P({\overline\sigma})$ for deriving,
through Equation \ref{eq:mass}, the expected distribution of galaxy stellar mass functions at early times. All that remains is to deduce its
amplitude/normalization.


\subsection{Deriving $\phi^{\ast}$: The Normalization of $P({\overline\sigma})$}
\label{sec:norm}

We have already derived the characteristic value of ${\overline\sigma}$ for a given halo mass, as well as the low-${\overline\sigma}$ power-law slope of the distribution of
$\sigbar$ at every $M_{h}$. We now turn to the normalization of $P({\overline\sigma})$.
The normalization consists of two pieces: (1) the of-order $\sim 10$ halos per Mpc$^{3}$ above the cooling limit at $z\sim 10$-$15$ \citep[e.g.]{tinker2008},
and (2) the total number of density fluctuations around these halos on very specific scales --- from the turnaround radius ($r_{\rm co}$) to the halo radius
($r_{h}$) --- again defined by the nonlinear power spectrum from which the collapsed halos are accreting. Let us first examine item (2).

A given galaxy seed embedded in a halo of mass $M_h$, at redshift $z_{\rm start}$, has a characteristic physical scale:
\begin{equation}
	r_h\sim \left[\frac{3M_h}{4\pi 178(1+z_{s\rm tart})^3\Omega_M\rho_{c}(1+\delta_{\rm LS})}\right]^{1/3},
\end{equation}
where $\rho_{c}$ is the comoving critical density. Smaller scales are, themselves, collapsed halos, so $r_h$ corresponds to a sharp break in the power-law
expectation for $P({\overline\sigma})$ at high values of ${\overline\sigma}$.

The halos are also decoupled from the surrounding linearly growing density field out to a comoving radius \citep[e.g.][]{evrard1992}:
\begin{equation}
	r_{\rm co}\sim \left[\frac{3M_h}{4\pi \Omega_M\rho_{c}(1+\delta_{\rm LS})}\right]^{1/3}.
\end{equation}

Density fluctuations on scales $r > r_{\rm co}$ define the background density on top of which the halos have contrast (i.e., within $r_{h}$) of $\Delta=178$
\citep{press1974}. While these scales are outside of the domain of $P({\overline\sigma})$, they are explicitly referenced in Equation \ref{eq:deltasigdep} as
$(1+\delta_{\rm LS})$, and thus define the characteristic ${\overline\sigma}^*$ for halos in a given over-density.

Density fluctuations on scales smaller than $r_{\rm co}$, down to $r_h$, are collapsed (sub)halos that have a
physical distribution described by the 1-halo term of the correlation function, $P(k)\sim k^{2\gamma-6}$, used in the previous section (Equation \ref{eq:pkgamma}).
These can be thought of as generating $(1+\delta_{\rm SS})$.

We therefore describe $P({\overline\sigma})$ as a power-law with lower and upper breaks defined by these two scales, $r\in[r_{h},r_{\rm co}]$. We model
the break at large ${\overline\sigma}$, or equivalently at small $r$, using an exponential cutoff as described by a \citet{schechter1976}
function.

To reiterate, $P({\overline\sigma})$ is more-or-less defined only between these two scales, $r_h$ and $r_{\rm co}$, with the latter radius defining the region that
decoupled from the rest of the universe, to grow (mildly) non-linearly according to prescriptions like \citet{peacock1996} and \citet{ma2000}.

Thus ---so long as a complete set of representative halos are considered --- $P({\overline\sigma})$ is normalized over the domain:
\begin{equation}
	\Delta \log r = \frac{1}{3}\log \left[178(1+z_{\rm start})^3\right],
\end{equation}
or, given Equation \ref{eq:r5}:
\begin{equation}
	\Delta \log {\overline\sigma} =\frac{x}{3} \log \left[178(1+z_{\rm start})^3\right].
\label{eq:sigspan}
\end{equation}

At early times --- $z_{\rm start}>10$ --- the range of small scale density fluctuations is thus approximately 9 dex in ${\overline\sigma}$
(assuming $x=5$ and initial accretion through streams with diameters uncorrelated with halo mass) --- equivalent to $9$ dex in stellar mass at
fixed time after the onset of star formation --- for a given ensemble represented by a particular ${\overline\sigma}^*$.
This expectation ought to be compared to the discussion in Section \ref{ref:mhalo}, in which the initial spectrum of halo masses results in a
$\simlt 1$ dex range of ${\overline\sigma}$ via the $M^{1/3}$ term in Equation \ref{eq:sigdep}.

Given an initial halo mass function with space densities of $\phi_h(M_h)$ of halos of mass $M_h$,
the normalization of the \citet{schechter1976}-like $P({\overline\sigma})$ can therefore be written
\begin{equation}
P({\overline\sigma}_{h}) = \frac{3\phi_h(M_h)}{\Gamma\left(-\frac{7}{5}+1, 10^{
-\frac{5}{3} \log \left[178(1+z_{\rm start})^3\right]}\right)}
\label{eq:norm}
\end{equation}
assuming $x=5$, with the factor of 3 due to the compression of objects into $1/3$ as many dex in ${\overline\sigma}$.
To give the approximate scale of this normalization, Equation \ref{eq:norm}
gives $P({\overline\sigma}_{h}) = 3\times 10^{-4}\phi_h(M_h)$ at $z\sim 10$.

\begin{figure*}[t]
\centerline{\includegraphics[width=0.95\hsize]{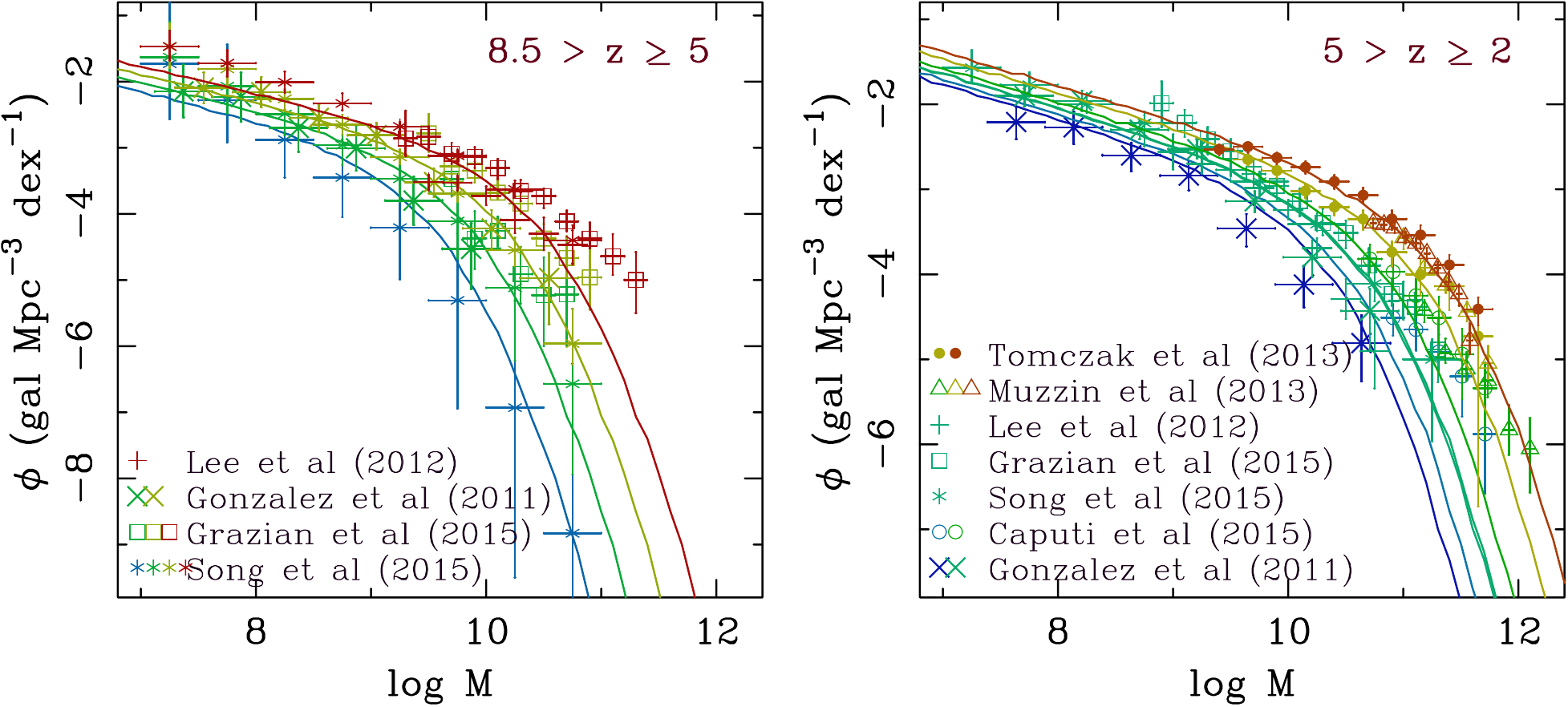}}
\caption{Observed stellar mass functions from $z=8$ to $z=2.5$
\citep{gonzalez2011,lee2012,tomczak2014,caputi2015,grazian2015,song2015}, all corrected to a \citet{salpeter1955} IMF.
The \cite{gonzalez2011} mass functions have been corrected systematically by offsets published by \cite{stark2013}. 
Assuming $z_{\rm start}=12.5$, we show theoretical stellar mass functions using the solid lines assuming
the accretion of baryons occurs along streams with mean diameters that are independent of halo mass.
While the true stellar mass functions should take into account a spectrum of large scale over- and under-densities that
will modulate ${\overline\sigma}^*$, as well as any intrinsic dispersion in $z_{\rm start}$, the simplest model shown here do
a fair job of reproducing the basic characteristics of published data.
\label{fig:mfevol}}
\end{figure*}

Summing up, at $z\sim 10$-$15$ there are $\sim 10$ dark matter halos per Mpc$^3$ that can support sustained star formation. These span
$\simlt 2.5$ dex in halo mass, adding $\simlt 1$ dex of scatter to the resulting stellar mass functions. However,
the broad range of matter densities surrounding these galaxy seeds produce a spread of $9$ dex in the abscissa of $P({\overline\sigma})$
when baryon accretion occurs along streams with mean diameters that are uncorrelated with halo mass.
And this spectrum of local matter densities produces a $9$ dex spread in stellar mass
until a time at which additional physics decelerates the {\it in situ\/} growth of high mass galaxies (i.e., halos residing in regions of high density at early times)
through the second term in Equation \ref{eq:d2mdt2}.

Even if the accretion was isotropic ($x=3$), the span in ${\overline\sigma}$ expected from the power spectrum of density fluctuations is more than 5 dex.
Recall that the range of halo mass above the cooling limit is only $\sim 2.5$ dex,
so that local density still remains significantly more important in defining the breadth of the stellar mass function (with halo mass providing some additional, meaningful broadening).
Regardless of the anisotropy of baryon accretion, the stellar mass function of galaxies should span $>7$ dex or so in mass at early times.

That the stellar mass functions of galaxies should span so many orders of magnitude, all the way back to the epochs at which sustained star
formation begins, is a prediction that will be easily testable using James Webb Space Telescope data.


\section{Theoretical Stellar Mass Functions and Their Evolution}
\label{sec:mfs}

\subsection{The First Few Gyr of Galaxy Growth}

\begin{figure}[h]
\centerline{\includegraphics[width=0.8\hsize]{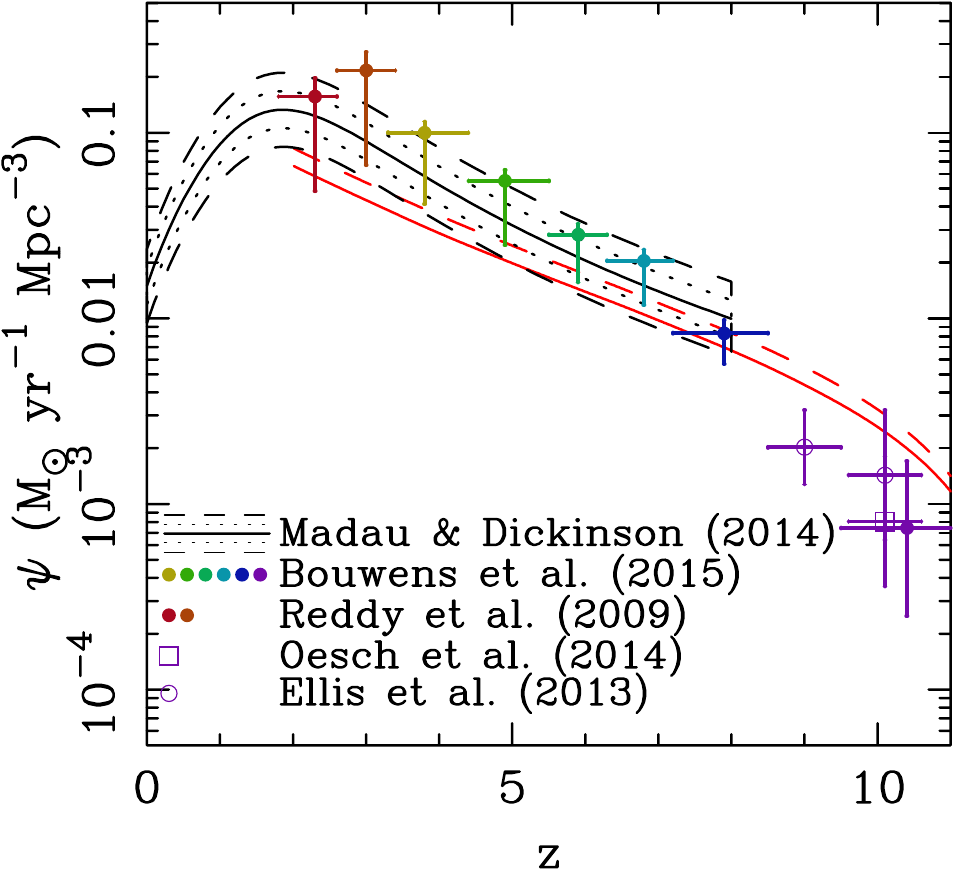}}
\caption{Assuming star formation begins at $z_{\rm start}=12.5$, we plot the theoretical
evolution of star formation rate density for the first 3 Gyr of the universe using the solid red line assuming galaxies accrete baryons
from streams with cross sections that are independent of halo mass. The dashed line shows the case for isotropic accretion.
The solid black line shows the functional
form of evolution from \citet{madau2014}, where we use dotted and dashed lines to illustrate potential uncertainties in
the SFRD evolution at levels of $\pm 0.1$ and $\pm 0.2$ dex. Published SFRD data from \cite{reddy2009,ellis2013,oesch2014,bouwens2015} are
shown using the colored points.
\label{fig:sfrdevol}}
\end{figure}

Figure \ref{fig:mfevol} shows the expected stellar mass functions predicted by our formalism (lines) superimposed on data from the literature
spanning $2\simlt z \simlt 8.5$ (points with error bars). We use the \citet{tinker2008} halo mass function at $z_{\rm start}=12.5$ down to
an atomic cooling limit at that redshift of $M_h>1.2\times 10^7 \Msun$. The theoretical mass functions are derived
for the case of $x=5$, stream accretion with diameters that are independent of halo mass, with ${\overline\sigma}^*$ drawn from Equation
\ref{eq:sigstar}\footnote{For completeness we show stellar mass functions for isotropic accretion (Case 1) and accretion through fixed
fractions of halo surface area (Case 2) in Appendix \ref{sec:app2}. Certainly the data at the highest redshifts may be consistent with such a scenario,
given basic uncertainties around key parameters, but the models shown in Figure \ref{fig:mfevol} provide better matches to galaxy data.}
is used explicitly, with ${\overline\sigma}^*$ from Equation \ref{eq:sigstar}. Figure \ref{fig:sfrdevol} shows the corresponding star formation rate densities.

Remarkably, with the {\it only\/} free parameter being a single mean starting redshift, $z_{\rm start}=12.5$,
our model predictions match the basic characteristics of published data at all epochs before the SFRD peaked \citep[e.g.][]{madau2014}; i.e. in the
era when the universe provided abundant fuel for stellar mass growth.
There is some flexibility given selection issues in the SFRD data, uncertainties in $M/L$ estimates from deriving stellar mass functions using rest-frame
UV data at the highest redshifts, and cosmic variance. But, overall, our basic picture of stellar mass growth as a stochastic process provides as good a match to published
data as can be hoped, given the extreme simplicity of the picture, and especially given the notable absence of astrophysical prescriptions for feedback
and star formation.

Using the mathematics of stochastic processes, we conclude that:
\begin{enumerate}
	\item the seeds of galaxies typically began growing at redshift $z_{\rm start}\sim 10$--15;
	\item in these seeds, star-formation began in proportion to the second derivative of baryon accretion at $z_{\rm start}$;
	\item those accretion rates were themselves determined by local matter density;
	\item {\it in situ} stellar mass growth then proceeded in manner consistent with stationarity, possibly implying quasi-steady state growth;
	\item those $\dot{M}$ states changed stochastically over time, but in ways that, for a given galaxy, were significantly correlated on the broad range of timescales as
implied by the power spectrum of density fluctuations on all scales ($H=1$). 
\end{enumerate}

With only collections of stellar masses and star formation rates for fair ensembles of galaxies at these early times, our results would suggest that little else can be said about
galaxy growth without more detailed observations on individual galaxies.

Our description of galaxy growth breaks the notion that there always exists a one-to-one mapping between halo mass and stellar mass.
At $z_{\rm start}$, or at inconsequentially small times after $z_{\rm start}$, galaxies with low stellar masses do not correspond to low mass
halos, but are better described as average halos with a low density of matter from which to accrete to build their halos and their stars.
Those halos start out initially with low rates of (halo and stellar mass) growth, rates of growth that persist on average for the ensemble of halos described by those conditions
(i.e. normal mass, low density). In this way, no tight stellar mass-to-halo mass relation should exist at these times.

Where (normal mass) halos have a low density of fuel (also having a low density of associated dark matter for accretion), they have ${\overline\sigma}$ values that are low. The
subsequent $t^2$ growth ensures that both stellar mass and dark matter mass remain jointly low for much of cosmic time. But those (normal mass) halos with a high density of
material for accretion (baryons and dark matter), have high values of ${\overline\sigma}$, thus comprising halo ensembles that will grow rapidly in both stellar mass and halo
mass.

As these ensembles evolve, the future stellar mass-halo mass relation emerges, and emerges rapidly, because $t^2$ growth diversifies the low- and high-density populations
quickly (e.g. Section \ref{sec:mgrowth}). In other words, even if all halos at $z\sim 10$-$15$ started star formation with the same halo mass, those with low values of
${\overline\sigma}$ would, on average, remain outpaced by those halos with high values of ${\overline\sigma}$. The low-${\overline\sigma}$ galaxies would, on average,
maintain low stellar and halo masses compared to those that had a greater abundance of local material and thus high values of ${\overline\sigma}$.

By, e.g., $z\sim 8$ or so, a well defined stellar mass-halo mass relation should be in place. Eventually ensembles with a high density of
fuel, characterized again by high values of ${\overline\sigma}$ and subsequent rapid growth, will reach end-states (e.g. exhaustion of fuel? high entropy? i.e. the second
term in Equation \ref{eq:d2mdt2} in Section \ref{sec:mfs} and Appendix C, below) more quickly than the halos that started out initially with a low density of fuel, being cursed
thereafter with low mean ensemble growth rates to stay at the lower end of the stellar mass function.

Putting this all together, it is the distribution of matter densities described by the one-halo term of the two-point correlation function that drives the wide range of stellar
masses spanned by the stellar mass function, and this wide range is frozen in place {\it for each initial ensemble $I$\/}, in effect for all subsequent cosmic time (until, e.g.,
fuel supplies run out in regions of different $\delta_{LS}$).

In this paradigm density, most notably density at $z_{\rm start}$, equals destiny.

\subsection{What Happens at Late Times?}
\label{sec:late}

Our formalism explicitly related stellar mass growth to baryon accretion, itself a process proportional to the dark matter accretion --- all at a single epoch, $z_{\rm start}$.
But if the constant of proportionality holds in some global sense over cosmic time, then the integral of $M\phi(M)$ at each redshift should be equal to the total mass of dark
matter involved in making galaxies, multiplied by $\epsilon_* f_b$. Interestingly, this calculation implies that all of the matter content of the universe has been involved in the
process of making galaxies at about $z\sim 2$. By this time the universe should no longer be able to sustain unabated stellar mass growth and the above description of the universe,
as a gigantic ensemble of uniformly efficient star factories must break.

Fortunately, the central limit theorem holds when a stochastic process is expected to be modified by long-term trends. In our case the radical redistribution of the fuel supply
and/or the injection of energy and/or entropy into that material will modify long-term expectations, each with potentially different mathematical characterizations. In Appendix
\ref{sec:app3}, we expand on the notion of entropy as related to the number of galaxy states available to the system but note that a more thorough examination of how the system
falters must be postponed for future papers.

No doubt the spectrum of large-scale density fluctuations also plays a crucial role, as $P({\overline\sigma})$
is shifted to larger values of ${\overline\sigma}$ in over-densities via the $(1+\delta_{\rm LS})^{5/3}$ term of Equation \ref{eq:deltasigdep}. Over-dense
regions of the universe reach their breaking points sooner, leading
the massive galaxies in those regions to their natural end states more quickly than those in lower density regions \citep[e.g.,][]{holmberg1965,dressler1980,abramson2016,morishita2016},
and leading to the hierarchical formation of galaxy groups in high density regions of the universe sooner than lower density regions \citep[e.g.,][]{wechsler2006}.

In fact, \citet[][see his Section VI]{dressler1980} had suggested that the Morphology-Density relation may have originated because high frequency (galaxy scale)
density perturbations were coupled to larger (cluster/group scale) density perturbations.
Here, we have derived just such a correlation between the values of ${\overline\sigma}$ that govern the rates of stochastic growth for ensembles of galaxies.
While we couched much of our work in Section \ref{sec:sigbar} in terms of baryonic acceleration, it must not be forgotten that ${\overline\sigma}$ is
also explicitly a measure of the RMS stochastic fluctuations in the star formation histories of galaxies. In other words, our derivations show directly that there can
be no separation between the evolution of galaxies and of their environments: low- and high-frequency perturbations are inextricably coupled, and both drive galaxy evolution.

Large and deep galaxy surveys at $z\lesssim 2$, such as CSI \citep{kelson2014b}, Z-FOURGE \citep{straatman2016}, GAMA \citep{driver2009}, and the SDSS \citep{york2000} will help
constrain how one writes the long-term expectations at later times by matching the detailed dependencies of the stellar mass function, as well as the scatter and slope in
the star-forming main sequence with galaxy mass and environment. With presumably some assumptions about how galaxy structure is related to phases of stellar mass growth---and
therefore the accumulation of angular momentum---perhaps progress can be made towards something approaching a first-principles treatment of galaxy evolution that reproduces the
many correlations of galaxy properties with mass and density over all time.

\section{Conclusions}
\label{sec:disc}

The central limit theorem allows for probability distributions and thus ensemble expectation values to be derived for any galaxy property that can be described as a
stochastic process. At the beginning of this endeavor, we wrote down {\it in situ\/} galaxy stellar mass growth as a stochastic process. We then explicitly related the form
of the expectation values for star formation rates to the accretion of baryons onto the early seeds of galaxies. Using this relation, we derived the evolution of the stellar
mass function and cosmic SFR density at times when fuel for SF was abundant: $2\lesssim z\lesssim 12$. Both reproduce existing data very well, especially considering that our model
has one---and only one---free parameter: $z_{\rm start}$, the redshift at which star formation begins.

We explored the dependencies of mean stellar mass growth on initial halo mass and on the initial spectrum of local matter densities, deriving the slope of the low mass end of the
stellar mass function, and without the need to invoke systematic changes in star formation efficiency or variable feedback. We identified a break at high mass in the stellar mass
function as arising from the smallest physical scale of density fluctuations in the nonlinear power spectrum, and derived a low-mass cutoff from the physical scale of turnaround
radius for halos at the onset of star formation. Together with the slope, these yielded the normalization of the stellar mass function in relation to the abundance of halos with
sufficient mass to support atomic cooling and initiate star formation. These things were neither expected, nor derived previously.

This work represents early efforts to derive an entirely new mathematical framework for exploring important aspects of galaxy formation and evolution, and without a heavy reliance
on the typical parameters that historically require tuning in simulations or semi-analytical models. As such, future derivations of chemical enrichment, two-halo clustering
statistics, and correlations of galaxy properties with the growth of structure, may then be significantly less dependent on assumptions that underpin previous galaxy evolution
models.

Of course inferring the physical manifestations of the stochastic growth we have posited requires both analytical and numerical
approaches to galaxy dynamics, which have been occurring now for decades
\citep[e.g.][]{toomre1972,vogelsberger2014,hopkins2014}, with exciting new analytical work on how stochastic effects drive
secular evolution in disks \citep[e.g.][]{fouvry2015a,fouvry2015b}.

The power spectrum of density fluctuations shapes the galaxy stellar mass function, as well as the rates at which ensembles of galaxies
evolve, with stochastic changes to galaxy states correlated on a broad range of timescales. Such correlations between stochastic events is often referred to as long-term memory, as
if whatever happens to them early in their lives remains important in the distant future. Some mathematicians argue that this phenomenon of long-term memory is merely the
imposition of large scale, long-term fluctuations in the global climate on the underlying intrinsic stochasticity \citep{samorodnitsky2006}. For galaxies this may be a distinction
without a difference: environment (large scale density fluctuations) acts as repositories of ``memory'' and actively drives climate change on long timescales. Simultaneously,
Hurst parameters approaching $H=1$ naturally arise out of positive stochastic flows \citep{koutsoyiannis2002}, bolstering work pointing to
the importance of cosmic flows in funneling baryons into accretion streams \citep[e.g.][]{dekel2009,tacchella2016}.

The specific details of inflow, outflow, or feedback, i.e. sub-grid physics, may indeed be important for understanding why galaxies look they way they do and have the chemical
enrichment histories that they do. However, in our view the detailed parameterizations of sub-grid processes are not particularly important for understanding the evolution of the
stellar mass content of galaxies, aside, possibly, from conditions of quasi-steady state, a mean value for star formation efficiency,\footnote{Stochastic variations around that
mean would be degenerate with the macro-physical variations that are subsumed in the stochastic process itself. No doubt, bulk observations of ensemble mass functions and star
formation rates are not likely to yield constraints on such subtle variations in the microphysics simply because of such degeneracies.} and correlations between changes to galaxy
states on an infinite range of timescales.

Crucially, our framework would suggest that super-galactic and galactic scale phenomena cannot be decoupled, but both depend fundamentally on aspects of the matter power spectrum
established at very early times. As such, stochastic growth {\it can and will\/} manifest in organized and coherent SFHs that depend on local environment.
Further, the permanent diversity of these SFHs will support the scatter in many scaling laws, including the SF main sequence and possibly the galaxy size--mass relation.
And given the hierarchy of halo assembly out of the fragmentary gravitational collapse of cosmological volumes, the stationary stochastic process already quite naturally provides a
framework with which to derive the accelerated evolution of some galaxy ensembles compared to others.

Put most simply:  the growth of galaxies for the first few Gyr of the universe could only proceed the way it did.
It is not sensitive to the detailed microphysics that complicate the fine tuning of
traditional astrophysical simulations, so long as stellar mass growth can be written as a stationary process to a sufficiently
high degree of accuracy. Again, the property of stationarity stems naturally from equilibrium frameworks of
the kind discussed by \cite{tinsley1978} and many since, in which star formation occurs in quasi-steady states \citep[c.f.][]{little2008}.
When fuel is plentiful, the addition of new sources of feedback are not particularly relevant so long as galaxies
are systems that, on average, figured out how to form stars without blowing themselves out or over-cooling.

We have gotten very far, from very little, in explaining evolutionary trends in galaxy star formation and stellar mass, using only the
expectation values of stochastic processes, Newtonian gravity, and star-formation beginning in proportion to a spectrum of baryon
accretion acceleration expected at $z\sim 10$--15. The flip-side, however, is that we have swept aside
the astrophysical questions that set $\epsilon$, the underlying physics that is involved, presumably, in keeping galaxies in quasi-steady state, at
least on some unknown timescale.

Ours, however, is a community deeply interested in such details, and we must continue searching for clues
to constrain the astrophysics of galaxy formation in whatever data we can. Unfortunately, the ensemble growth of stellar
mass in the universe appears not to be suitable for this endeavor.

What this ultimately means for astronomers interested in the astrophysical processes that change galaxies over time is that
ensemble measurements --- at as many epochs as one likes --- are unhelpful at best, and detrimental to understanding at
worst. For it has been the slow, steady apparent evolution of the properties of cosmic ensembles that has lulled us into forgetting how
dynamic the cosmos really can be. Many of
the things we need to learn about how galaxies form, how they can change rapidly, and how they evolve secularly, simply cannot be learned
from comparing different samples at different redshifts with each other, whether
selected by mass or by comoving number density, even if systematic corrections to these methodologies are made \citep{torrey2016}.
The disconnect between what observers glean from the statistics of
galaxy evolution and the obvious dynamism in hydrodynamical $n$-body simulations has so far not been bridged by interpreting the slow
evolution of ensembles as representative of its constituents, and patching such approaches only serves to mask the underlying diversity
in growth and assembly.

Fortunately, mathematical tools exist for
properly treating that dynamism in the steady evolution of the cosmic ensembles of galaxies.
The mathematics of stochastic processes is powerful, and provides a means by which to model the distribution of
galaxy growth trajectories. With that knowledge, that is the (full) set of underlying histories that go into the making of
a set of observations, the community should then be able to utilize
detailed, potentially very deep, observations on (presumably small) samples, with
the explicit goals of illuminating the side effects of cosmic dynamism, by specifically probing the warping of disks, measuring the heating of velocity
ellipsoids with galaxy position, tracing the dependencies of internal kinematics on stellar population ages, all together, and in conjunction
with specific, representative, and properly-motivated growth histories. It will be such detailed observations and sophisticated modeling of physical processes that will allow us to
carefully disentangle the processes that enable galaxies to dance between blowing themselves out or making it for the long haul.


\acknowledgements

The first author thanks his family for putting up with particularly intense crankiness during this endeavor. Gratitude for
the first author's awesome, most bodacious collaborators, for they were patient with the fits and starts and hiccups along
the way. Appreciation is expressed in advance to B.~Weiner for Twitter comments. A final shout-out to S.~.C.~Trager is also
in order for the foggy, late-night 1990's graduate school discussions about how there must exist multiple paths to making
similar kinds of galaxies.



\appendix

\section{A. A Characteristic Value for $P({\overline\sigma})$: Cases (1) and (2)}
\label{sec:app1}

In the case of isotropic accretion (Case 1, $C=1$), or at
least accretion through fixed fractional covering areas (Case 2), the spectrum of $P({\overline\sigma})$ would have a characteristic value
\begin{eqnarray}
{\overline\sigma}^* =
\left(\frac{\epsilon_{\rm ff}}{0.015}\right)
\left(\frac{\gamma}{2}\right)
\left(\frac{f_b}{0.16}\right)
\left(\frac{C}{0.3}\right)
\left(\frac{1+z_{\rm start}}{1+12.5}\right)^3
\left(\frac{M^*_{h,z_{\rm start}}}{1\times 10^9 \Msun}\right)
\left(2.9 \times 10^{-7} \Msun/\text{yr}^2\right)\qquad
\label{eq:sigstariso}
\end{eqnarray}

\section{B. Stellar Mass Functions for Cases of Isotropic Accretion}
\label{sec:app2}

\begin{figure*}[h]
\centerline{\includegraphics[width=0.95\hsize]{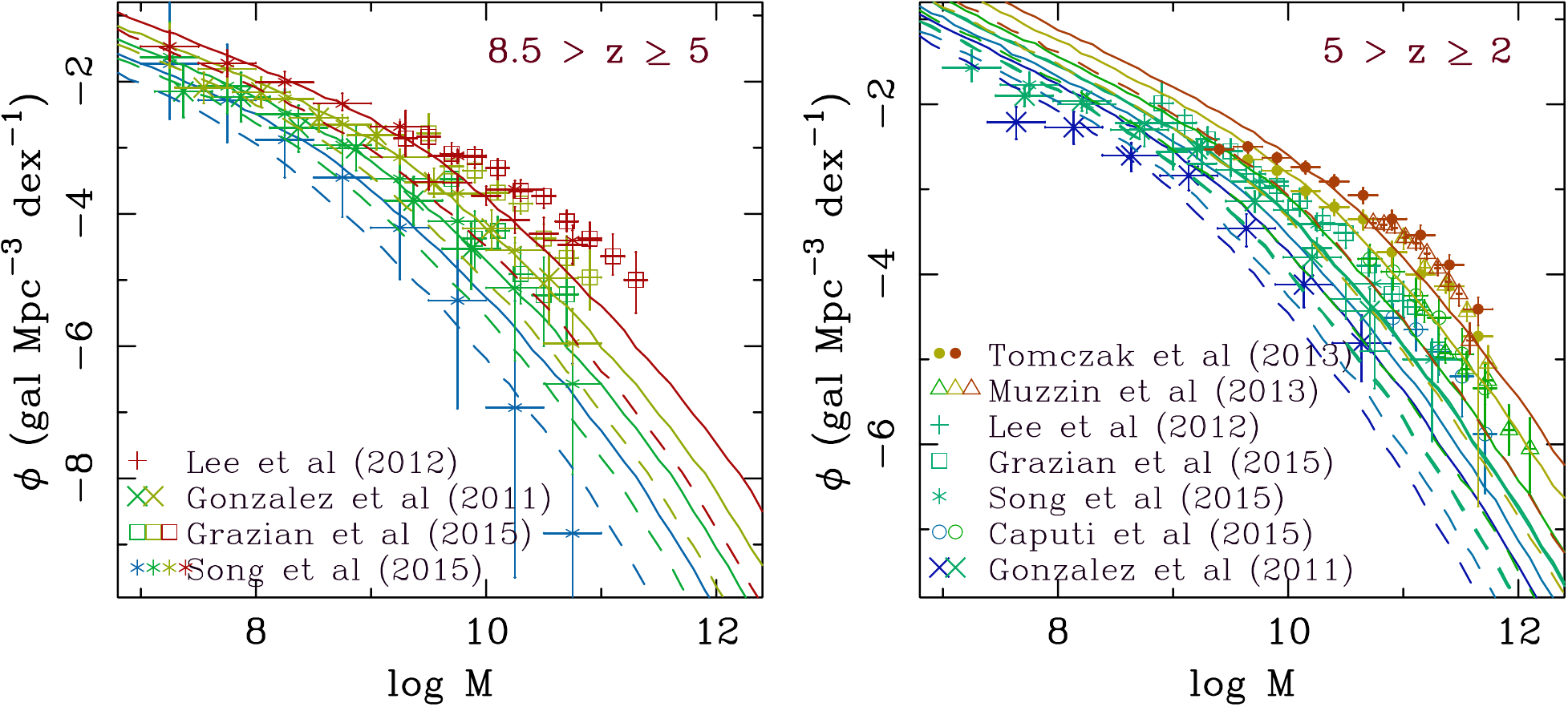}}
\caption{Same as in Figure \ref{fig:mfevol} but assuming accretion occurs isotropically, $C=1$, solid line; or through streams
with covering fraction of $C=0.3$, dashed line.
\label{fig:mfevoliso}}
\end{figure*}

Our picture is moderately sensitive to the adopted form of the accretion anisotropy.
Figure \ref{fig:mfevoliso} shows the model stellar mass functions for the case where the accretion occurs through a fixed fraction of
halo surface area. In general, these do not match the shape of published stellar mass functions for high redshift galaxies and we
attribute this to the nature of the accretion streams with respect to the halos initiating star formation.

\section{C.Considering Entropy}
\label{sec:app3}

We here speculate on another aspect of the second derivative of the mean stellar mass growth of galaxy ensembles.
The second term of Equation \ref{eq:d2mdt2} is not uniquely described by breakdowns in the available fueling of star formation,
as it is also proportional to the change in the log of the sum of the variances to previous changes to galaxy states. In other words,
the system's continued rate of evolution is dependent on the number of possible galaxy states, highly reminiscent of the notion of entropy
for an ensemble in a statistical mechanical sense.

Using \cite{shannon1948} let us define an entropy for ensemble $I$ at time $t$:
\begin{equation}
\mathcal{H}_{I,t} = \ln \sqrt{2\pi e}{\rm Sig\/}[{\dot M}_{I,t}]
\label{eq:shannonH}
\end{equation}
then
\begin{equation}
\frac{d\mathcal{H}_{I,t}}{dt} = \frac{d\ln{\overline\sigma}_{I,t}}{dt} + \frac{1}{t}
\label{eq:dHdt}
\end{equation}
Substituting, Equation \ref{eq:sigeq} becomes
\begin{equation}
{\rm E\/}\left[ \frac{d^2M}{dt^2}\right]_I =
\frac{\overline{\sigma}_{I,t}}{\sqrt {2\pi }}t \frac{d \mathcal{H}_{I,t}}{dt}
\label{eq:entropy}
\end{equation}
in which \cite{shannon1948}'s entropy, no longer referring to ensembles of messages or strings of information, instead refers to the number of states and/or to
the number of state changes accessible to galaxy growth trajectories. When ${\overline\sigma}$ is independent of time, $td\mathcal{H}_{I,t}/dt=1$
and Equation \ref{eq:entropy} reduces to Equation \ref{eq:sigeq}.

At some point, as the universe's mass distribution becomes ever
more disrupted by the processes of gravitational collapse and dynamics,
particular ensembles may
reach a point where $\mathcal{H}_{I,t}$ no longer increases as quickly as $\ln t$ -- which is another way of stating that the number
of accessible galaxy states encompassed by ${\overline\sigma}$ may become restricted.

Such an approach may lead us to find matter density thresholds (in $\delta_{LS}$) that restrict the availability of particular galaxy states
in high-entropy regions of the universe, though this is unproven speculation. Nonetheless, if such
parameterizations were to provide the best fit to late-time data on galaxy ensembles, the concordance with
\citet{voit2015a,voit2015b} would be remarkable. Those authors determined that
thermodynamic entropy plays a large role in diminishing on-going star formation in massive systems,
restricting the precipitation of cold gas clouds. Note, too, that this restriction on galaxy states and star formation
occurs relative to the ({\it ex situ\/}) stellar mass growth that continues via gravitational assembly.

In practice there may be several reasons for $\mathcal{H}(t)$ to, at some point, grow more slowly than $\ln t$. The long-term stellar mass
growth may simply decelerate owing to diminished availability of material. In principle this may not modify the underlying (Gaussian or
fractional Gaussian) noise but may merely change long-term expectations in the mean, shifting the
centroid of the probability distribution produced by the central limit theorem. Or there may be physical processes, in
specific later-time environments, for example, that restrict the possible states or state changes for galaxies.
Making these distinctions will require specific data to interpret the physical manifestations of the mathematics.
Fortunately for us, and the reader, the dual roles of finite fuel supplies and entropy evolution are
well beyond the scope of this initial work and are reserved for later papers.

\section{D. Further Interpretations of Stationarity}
\label{sec:app4}
Stationarity naturally arises out of discussions of steady state equilibrium (see references given earlier), but the condition of
steady-state is not technically required in order for us to represent stellar mass growth as a stationary stochastic process. The Queueing
of material for star formation implies stationarity at least on cosmologically short timescales through Little's Law
\cite[see][]{little2008,whitt1991}, even when systems are in unsteady states.

Little's Law is essentially a restatement of conservation of mass for Queueing systems. In our case, we have the stochastic arrival of baryons into a system providing a
``service:'' the conversion of baryons to stars. That service, proceeding at a roughly fixed rate \cite[at least in relation to a free-fall time][]{krumholz2005}, is a bottleneck in
the process of {\it in situ\/} stellar mass growth in that it is a step intermediate between baryon accretion and star formation. So long as all baryons slated for that service are
converted to stars in the bookkeeping (i.e. the mean efficiency remains unchanged with time), then Little's Law holds, and we can adopt stationarity without the explicit
requirement of (quasi-)steady states.

\section{E. What are the Halos Doing?}
\label{sec:app5}

One might also consider whether published analytical forms for dark matter accretion rates of halos \citep[e.g.][]{mcbride2009,fakhouri2010} might inform us of the appropriate
second derivatives. Surely the time derivative of such parameterizations, at $z_{\rm start}$, for a given halo mass should be all that is required. Unfortunately (or fortunately,
depending on your perspective), \citet[][in prep]{kelson2017} show that the growth trajectories implied by published accretion rates have derivatives consistent with dark matter
halo growth as its own stationary stochastic process that begins at earlier epochs, when specific modes of the power spectrum reach $\Delta^2(k)\sim 1$ and transition to nonlinear
growth.

Put more simply, the mean rates of growth for ensembles of halos were defined by the density field at much earlier times than the epochs we are concerned with here, the epochs when
(sustained) star formation begins.  The mean growth of dark matter halos track Equation \ref{eq:mass1} but the clock for the ensemble defined by a given wavenumber is not
synchronized with the clock that starts when star formation begins. And while delving into the nonlinear growth of halos at early times is beyond the scope of this paper, it is the
focus of \citet[][in prep]{kelson2017} along with a demonstration of the extent to which the equations worked through in this paper are valid for the stochastic process of dark
matter accretion.

\end{document}